\RequirePackage{fix-cm}

\documentclass[twocolumn,epjc3]{svjour3}  
\pdfoutput=1 
\usepackage{atlasphysics} 
\usepackage{subfigure}

\usepackage{rotating}
\usepackage{graphicx}
\usepackage{bigstrut}
\usepackage[hyperindex,breaklinks]{hyperref} 

\usepackage{multirow}
\usepackage{cite}
\usepackage[font=small,labelfont=bf, labelsep=quad]{caption}
\usepackage{amsmath}
\usepackage{tikz}
\usetikzlibrary{arrows,shapes}
\tikzstyle{every picture}+=[remember picture]
\usepackage[overlay]{textpos}
\newenvironment{changemargin}[2]{\begin{list}{}{%
\setlength{\topsep}{0pt}%
\setlength{\leftmargin}{0pt}%
\setlength{\rightmargin}{0pt}%
\setlength{\listparindent}{\parindent}%
\setlength{\itemindent}{\parindent}%
\setlength{\parsep}{0pt plus 1pt}%
\addtolength{\leftmargin}{#1}%
\addtolength{\rightmargin}{#2}%
}\item }{\end{list}}

\usepackage[switch]{lineno}   
\newcommand{\minitab}[2][l]{\begin{tabular}{#1}#2\end{tabular}}
\smartqed  

\newcommand{\Wheltitle}{Measurement of the polarisation of \Wboson\ bosons produced with large
  transverse momentum in $pp$ collisions at $\sqrt{s}$ = 7\ \TeV\ with the ATLAS experiment}

\journalname{Eur. Phys. J. C}

\makeatletter
\g@addto@macro\bfseries{\boldmath}
\makeatother

\usepackage{preprintcover} 
\PreprintCoverPaperTitle{\Wheltitle}  
\PreprintIdNumber{CERN-PH-EP-2012-016}  
\PreprintCoverAbstract{
This paper describes an analysis of the angular distribution of \Wen\ and \Wmn\ decays, using data from $pp$ collisions at $\sqrt{s}$ = 
7 \TeV\ recorded with the ATLAS detector at the LHC in 2010, corresponding to an integrated luminosity of about 35 \ipb. Using the 
decay lepton transverse momentum and the \MTE, the \Wboson\ decay angular distribution projected onto the transverse plane is obtained and 
analysed in terms of helicity fractions \fl, \fL\ and \fR\ over two ranges of \Wboson\ transverse momentum (\ptw): 35 $<$ \ptw\ $<$ 
50 \GeV\ and \ptw\ $>$ 50 \GeV. Good agreement is found with theoretical predictions. For \ptw\ $>$ 50 \GeV, the values of \fl\ and 
\fLmR, averaged over charge and lepton flavour, are measured to be : \fl\ = 0.127 $\pm$ 0.030 $\pm$ 0.108 and \fLmR\ = 0.252 $\pm$ 
0.017 $\pm$ 0.030, where the first uncertainties are statistical, and the second include all systematic effects.
}  
\PreprintJournalName{Eur. Phys. J. C}  

\begin{document}

\title{\Wheltitle}

\author{The ATLAS Collaboration}
\institute{CERN, 1211 Geneva 23, Switzerland}

\date{Received: date / Accepted: date }

\maketitle

\begin{abstract}
This paper describes an analysis of the angular distribution of \Wen\ and \Wmn\ decays, using data from $pp$ collisions at $\sqrt{s}$ = 
7 \TeV\ recorded with the ATLAS detector at the LHC in 2010, corresponding to an integrated luminosity of about 35 \ipb. Using the 
decay lepton transverse momentum and the \MTE, the \Wboson\ decay angular distribution projected onto the transverse plane is obtained and 
analysed in terms of helicity fractions \fl, \fL\ and \fR\ over two ranges of \Wboson\ transverse momentum (\ptw): 35 $<$ \ptw\ $<$ 
50 \GeV\ and \ptw\ $>$ 50 \GeV. Good agreement is found with theoretical predictions. For \ptw\ $>$ 50 \GeV, the values of \fl\ and 
\fLmR, averaged over charge and lepton flavour, are measured to be : \fl\ = 0.127 $\pm$ 0.030 $\pm$ 0.108 and \fLmR\ = 0.252 $\pm$ 
0.017 $\pm$ 0.030, where the first uncertainties are statistical, and the second include all systematic effects.
\end{abstract}

\section{Introduction}
\label{sect:intro}
This paper describes a measurement with the ATLAS detector of the polarisation of \Wbosons\ with transverse momenta greater 
than 35 \GeV, using the electron and muon decay modes, in data recorded at 7 \TeV\ centre-of-mass energy, with a total 
integrated luminosity of about 35 \ipb. The results are compared with theoretical predictions from \mcnlo\ \cite{Frixione2002ik} 
and \pow\ \cite{Nason2004rx,Frixione2007vw,Alioli2010xd,Alioli2008gx}. 

The paper is organised as follows. Section\ \ref{sect:theor} describes the theoretical framework of this analysis. Section\ 
\ref{sect:atlas} reviews the relevant components of the ATLAS detector, the data, the corresponding Monte Carlo simulated data sets, and 
the event selection. The estimation of backgrounds after this selection is explained in Section\ \ref{sect:normalisatiom}, and the comparison 
of data and Monte Carlo simulations for the most relevant variable (\ctdd) is given in Section\ \ref{sect:datamccomp}. The construction of 
helicity templates and its validation using Monte Carlo samples is described in Section\ \ref{sect:templatesandclosure}, while the uncorrected 
results are given in Section\ \ref{sect:fitresults}. The systematic uncertainties associated with the fitting procedure are discussed in 
Section\ \ref{sect:systematic} and the final results, corrected for reconstruction effects, are given in Section\ \ref{sect:final}. Section\ 
\ref{sect:conclusions} is devoted to the conclusions.

\section{Theoretical framework and analysis procedure}
\label{sect:theor}
Measuring the polarisation of particles is crucial for understanding their production mechanisms.\par
At hadron colliders, \Wbosons\ with small transverse momentum are mainly produced through the leading order electroweak processes
\begin{linenomath*} $$u\bar{d}\ra \Wp \qquad \mbox{ and } \qquad d\bar{u}\ra \Wm$$  \end{linenomath*} At 
the LHC the quarks generally carry a larger fraction of the momentum of the initial-state protons than the antiquarks. This causes 
the  \Wbosons\ to be boosted in the direction of the initial quark. In the massless quark approximation, the quark must be left-handed 
and the antiquark right-handed. As a result the \Wbosons\ with large rapidity (\yw) are purely left-handed.\par
For more centrally produced \Wbosons, there is an increasing probability that the antiquark carries a larger momentum fraction than 
the quark, so the helicity state of the \Wbosons\ becomes a mixture of left- and right-handed states whose proportions are respectively 
described with fractions \fL\ and \fR.\par

For \Wboson\ bosons with large transverse momentum, three main processes contribute (taking the \Wp\ as example): \begin{linenomath*} 
$$ug\ra \Wp d \;\; \mbox{, }\;\; u\bar{d}\ra \Wp g \;\; \mbox{ and } \;\; g\bar{d}\ra \Wp \bar{u}$$\end{linenomath*} Given the vector 
nature of the gluon, present in all three reactions, the simple argument used at low \ptw\ no longer applies. Predictions require 
detailed helicity state calculations. Leading-order (LO) and next-to-leading-order (NLO) QCD predictions have been available for 
$p\bar{p}$ interactions for some time\ \cite{Mirkes1992hu} and more recently for proton-proton interactions\ \cite{Bern2011ie}. At high 
transverse momenta more complex production mechanisms contribute, and polarisation in longitudinal states is also possible (the proportion 
of longitudinal \Wbosons\ is hereafter described by \fl). This state is particularly interesting as it is directly connected to the massive 
character of the gauge bosons.  

\subsection{Theoretical framework}
\label{sect:theorinput}
The general form for inclusive \Wboson\ production followed by its leptonic decay can be written 
as\ \cite{Mirkes1992hu}:
\begin{changemargin}{-0.025\textwidth}{0cm}
  \begin{eqnarray}
    \label{eqn:xsectot}
    \frac{d \sigma}{d(\ptw)^2 d\yw d\cos \theta d \phi} &=& \frac{3}{16 \pi}\frac{d \sigma^{u}}{d(\ptw)^2d\yw} \times \big[(1+\cos^{2} \theta) \nonumber\\
    & + &\frac{1}{2}A_{0}(1-3\cos^{2}\theta) + A_{1}\sin 2\theta \cos\phi \nonumber \\
    &+& \frac{1}{2}A_{2} \sin^2 \theta \cos2\phi + A_3 \sin\theta \cos\phi  \nonumber \\
    &+& A_4 \cos\theta + A_5 \sin^2\theta \sin2\phi \nonumber \\
    &+& A_6 \sin 2\theta \sin\phi + A_7 \sin\theta \sin\phi\big]
  \end{eqnarray}
\end{changemargin}
where $\sigma^u$ is the unpolarised cross-section and $\phi$ and $\theta$ are the azimuthal and polar angles of the charged lepton 
in a given \Wboson\ rest frame. The $A_i$ coefficients are functions of \ptw\ and \yw\ and depend on the parton distribution functions 
(PDFs). For $\ptw\rightarrow0$ all reference frames used in Refs.\ \cite{Mirkes1992hu,Korner1990im,Collins1977iv,Lam1978pu,Mirkes1994eb,
Berger2007jw,Bern2011ie} become identical, with the $z$-axis directed along the beam axis. In these conditions the dependence on $\phi$ 
disappears and only the term with $(1+\cos^{2} \theta)$ and the terms proportional to $A_{0}$ and $A_{4}$ remain.\par
The $A_0$ to $A_4$ coefficients in \Eqn{xsectot} receive contributions from QCD at leading and higher orders, while $A_5$ to $A_7$ appear 
only at next-to-leading order. Their expression as a function of \ptw\ and \yw\ depends on the reference frame used for the 
calculation.\par
Several papers have been published to discuss and predict these coefficients, first for $p\bar{p}$ colliders\ \cite{Mirkes1992hu,Korner1990im,
Collins1977iv,Lam1978pu,Mirkes1994eb,Berger2007jw} and more recently for the LHC\ \cite{Bern2011ie}. While at $p\bar{p}$ colliders, 
because of CP invariance, the $A_i$ coefficients are either equal ($A_0$, $A_2$, $A_3$, $A_5$, $A_7$) or opposite ($A_1$, $A_4$, $A_6$) 
for \Wp\ and \Wm\ production, there is no such simple relationship at $pp$ colliders. However it has been observed\ \cite{Bern2011ie} that 
$A_3$ and $A_4$ change sign between \Wp\ and \Wm, while the other coefficients ($A_0$, $A_1$, $A_2$, $A_5$, $A_6$, $A_7$) do not and are 
similar in magnitude between \Wp\  and \Wm. In all cases, the pure NLO coefficients ($A_5$ to $A_7$) are small. They are neglected in this analysis.\par
Experimental measurements have been reported from the Tevatron by CDF\ \cite{CDF2005angle}, from HERA by H1\ \cite{HERA} and recently from 
the LHC by CMS\ \cite{Chatrchyan2011ig}. 

\subsection{Helicity fractions}
\label{sect:polafrac}

\begin{figure}[h!]
  \begin{changemargin}{-0.025\textwidth}{0cm}
    \begin{center}
      \includegraphics[width=0.52\textwidth]{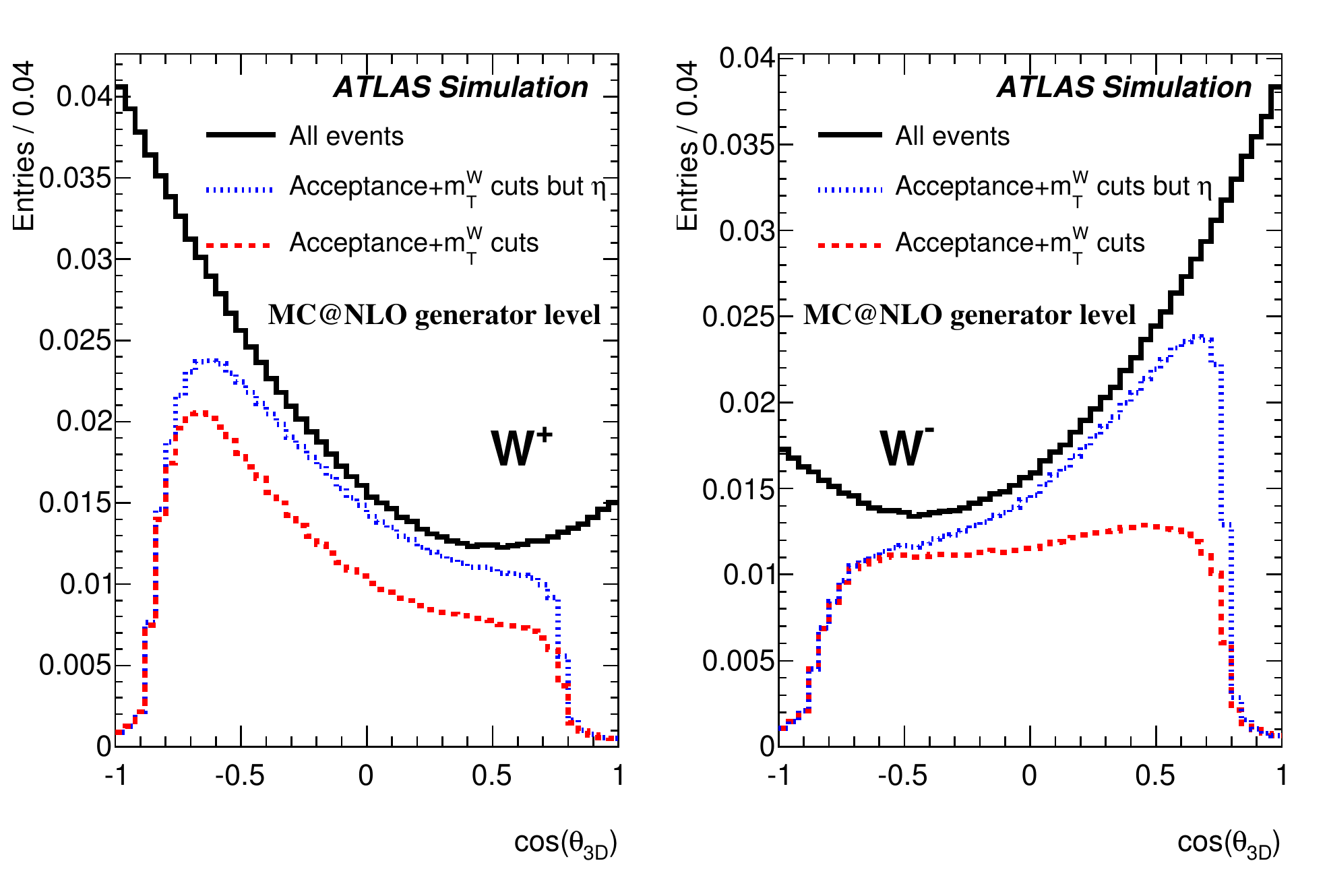}
    \end{center}
  \end{changemargin}
  \caption{Cosine of the helicity angle of the lepton from \Wboson\ decay at generator-level for positive charge (left) and negative charge 
    (right). Solid lines are without selection, dashed lines are after all acceptance plus \trm\ cuts except the \etal\ cuts and dotted lines 
    are after all acceptance plus \trm\ cuts. ``All events'' distributions are normalised to unity.\label{fig:Trcosine}}
\end{figure}

Helicity is normally measured by analysing the distribution of the cosine of the helicity angle ($\theta_{\rm 3D}$ in the following), defined 
as the angle between the direction of the \Wboson\ in the laboratory frame and the direction of the decay charged lepton in the \Wboson\ 
rest frame. The distribution of this angle as generated by \mcnlo\ is shown in \Fig{Trcosine} without phase space restriction, as well as 
with the acceptance (\ptl, \etal\ and \ptn)\footnote{ATLAS uses a right-handed coordinate system with its origin at the nominal interaction 
point (IP) in the centre of the detector and the $z$-axis along the beam pipe. The $x$-axis points from the IP to the centre of the LHC ring, 
and the $y$-axis points upward. Cylindrical coordinates $(r,\phi)$ are used in the transverse plane, $\phi$ being the azimuthal angle around 
the beam pipe. The pseudorapidity is defined in terms of the polar angle $\theta$ as $\eta=-\ln\tan(\theta/2)$.} and \Wboson\ transverse mass 
\trm\ cuts (where \trm=$\sqrt{ 2( \ptl \ptn - \vptl \cdot \vptn )}$), described in Section\ \ref{sect:selection}.\par

The differential cross-section in the helicity frame\footnote{The helicity frame is the \Wboson\ rest frame with the $z$-axis along the \Wboson\ 
laboratory direction of flight and the $x$-axis in the event plane, in the hemisphere opposite to the recoil system.} is expressed by using 
$\theta_{\rm 3D}$ and $\phi_{\rm 3D}$ in \Eqn{xsectot}. Integrated over  \yw\ and $\phi_{\rm 3D}$, \Eqn{xsectot} then takes the form:
\begin{eqnarray}
  \label{eqn:Aioverphihel}
  \frac{1}{\sigma}\frac{d\sigma}{d\ctd} =  \frac{3}{8} [(1+\cct_{\rm 3D})& + &A_0\frac{1}{2}(1-3\cct_{\rm 3D}) \nonumber\\
  & + &A_4\ctd].
\end{eqnarray}

Comparing \Eqn{Aioverphihel} to the standard form\ \cite{Ellis1991qj} using helicity fractions: 
\begin{eqnarray}
  \label{eqn:fracdef}
  \frac{1}{\sigma}\frac{d\sigma}{d\ctd} = \frac{3}{8}\fL (1\mp\ctd)^2 & + & \frac{3}{8}\fR (1\pm\ctd)^2\nonumber\\
  & + & \frac{3}{4}\fl \sct_{\rm 3D}
\end{eqnarray}
yields the relations between the $A_i$ coefficients and the helicity fractions:
\begin{linenomath*}\begin{alignat}{2}
  \label{eqn:fracAi}
  \fL(\yw,\ptw)  & = &&\, \frac{1}{4} (2-A_0(\yw,\ptw)\mp A_4(\yw,\ptw)) \nonumber \\
  \fR(\yw,\ptw)  & = &&\, \frac{1}{4} (2-A_0(\yw,\ptw)\pm A_4(\yw,\ptw)) \nonumber \\
  \fl(\yw,\ptw)  & = &&\, \frac{1}{2} A_0(\yw,\ptw) 
\end{alignat}\end{linenomath*}
where the upper (lower) sign corresponds to \Wp\ (\Wm) boson production respectively. It is interesting to notice that the difference 
between the left- and right-handed fraction is proportional to $A_4$ only, as:
\begin{linenomath*}\begin{equation} 
\label{eqn:a4rel}
\hfill \fLmR = \mp \frac{A_4}{2}. \hfill
\end{equation}\end{linenomath*}

From general considerations, the longitudinal helicity fraction \fl\ is expected to vanish for $\ptw\rightarrow 0$ as well as for $\ptw
\rightarrow\infty$, with a maximum expected around  45 \GeV\ \cite{Bern2011ie}.\par

\subsection{Analysis principle and variable definitions}
\label{sect:analyprinciple}

When analysing data, a major difficulty arises from the incomplete knowledge of the neutrino momentum. The large angular coverage of the 
ATLAS detector enables measurement of the \MTE, which can be identified with the transverse momentum of the neutrino. The longitudinal 
momentum can be obtained thro\-ugh the \Wboson\ mass constraint. However, solving the corresponding equation leads to two solutions, between 
which it is not possible to choose in an efficient way. The approach taken in this analysis is to work in the transverse plane only, using the 
``transverse helicity'' angle $\theta_{\rm 2D}$ defined by: 
\begin{linenomath*}\begin{equation} 
  \hfill \ctdd=\frac{\vptls \cdot \vptw}{|\vptls| \; |\vptw|}\,,\hfill
\end{equation}    \end{linenomath*}
where \vptls\ is the transverse momentum of the lepton in the transverse \Wboson\ rest frame and \vptw\ is the transverse momentum of the 
\Wboson\ boson in the laboratory frame. The angle $\theta_{\rm 2D}$ is a two dimensional projection of the helicity angle $\theta_{\rm 3D}$. 
Its determination uses only fully measurable quantities, defined in the transverse plane. Its use is limited to sizeable values of \ptw, 
which corresponds to the physics addressed in this work.\par 
The correlations between \ctdd\ and \ctd\ for events where \ptw\ $>$ 50 \GeV\ are represented in \Figs[ctdres1]{ctdres2} for positive and 
negative leptons respectively. This information is obtained using a sample of events simulated with \mcnlo\ after applying acceptance and 
\trm\ cuts.\par
The enhancement near $-$1 for positive leptons reflects that the maximum of the left-handed part of the decay distribution 
(first term in \Eqn{fracdef}) falls within detector acceptance, as opposed to the case of negative leptons where the maximum (near +1) falls 
largely beyond the \etal\ acceptance, resulting in a more ``symmetric'' distribution between forward and backward hemispheres. This effect is also 
seen in \Fig{Trcosine} when comparing \ctd\ distributions at generator-level, before and after the lepton pseudorapidity cut.\par 
\begin{figure}[h!]
  \begin{changemargin}{-0.05\textwidth}{-0.05\textwidth}
    \centering
    \subfigure[$\ell^+$]{
      \includegraphics[width=0.27\textwidth]{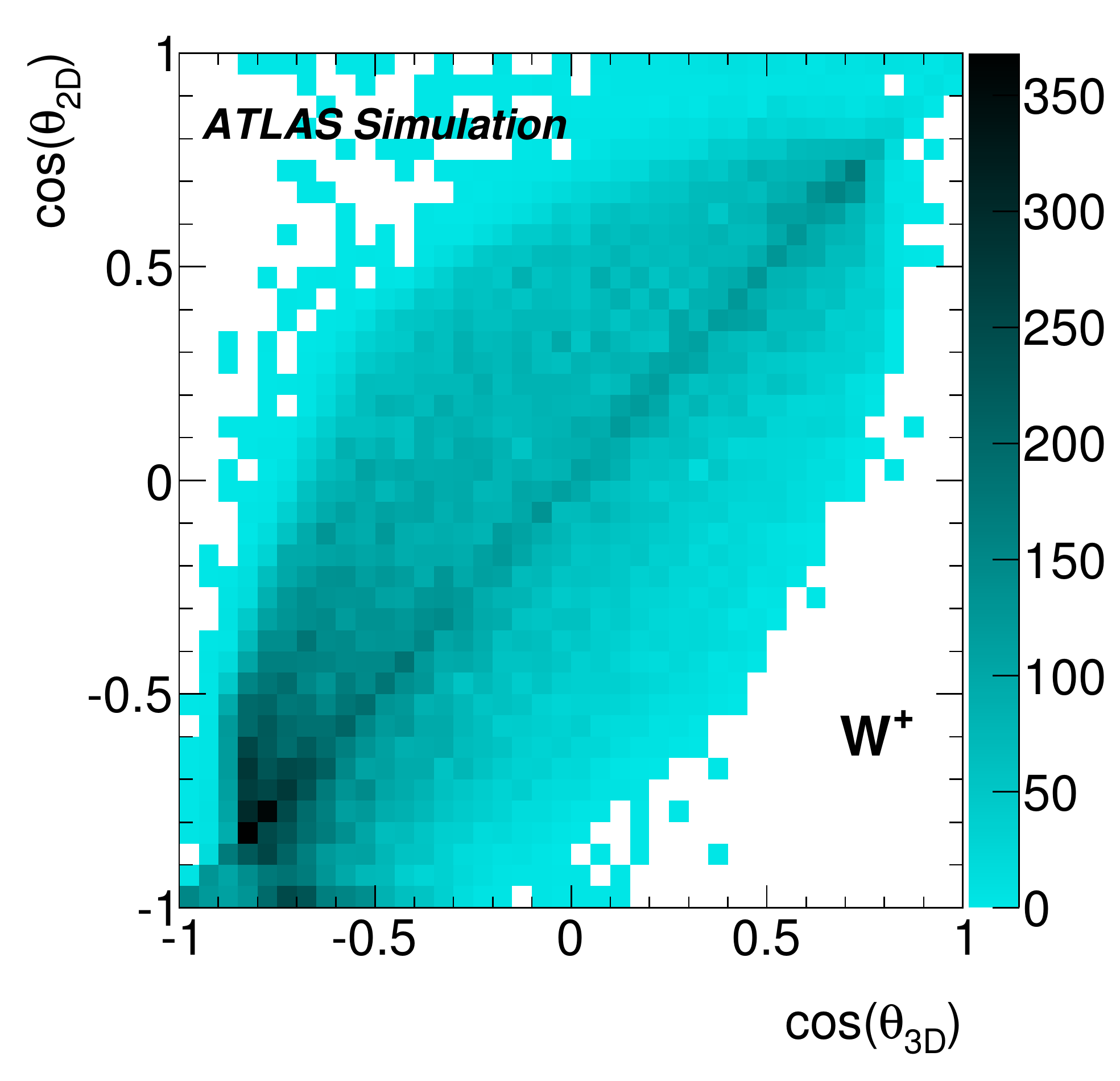}
      \label{fig:ctdres1}
    }\hspace*{-0.4cm}
    \subfigure[$\ell^-$]{
      \includegraphics[width=0.27\textwidth]{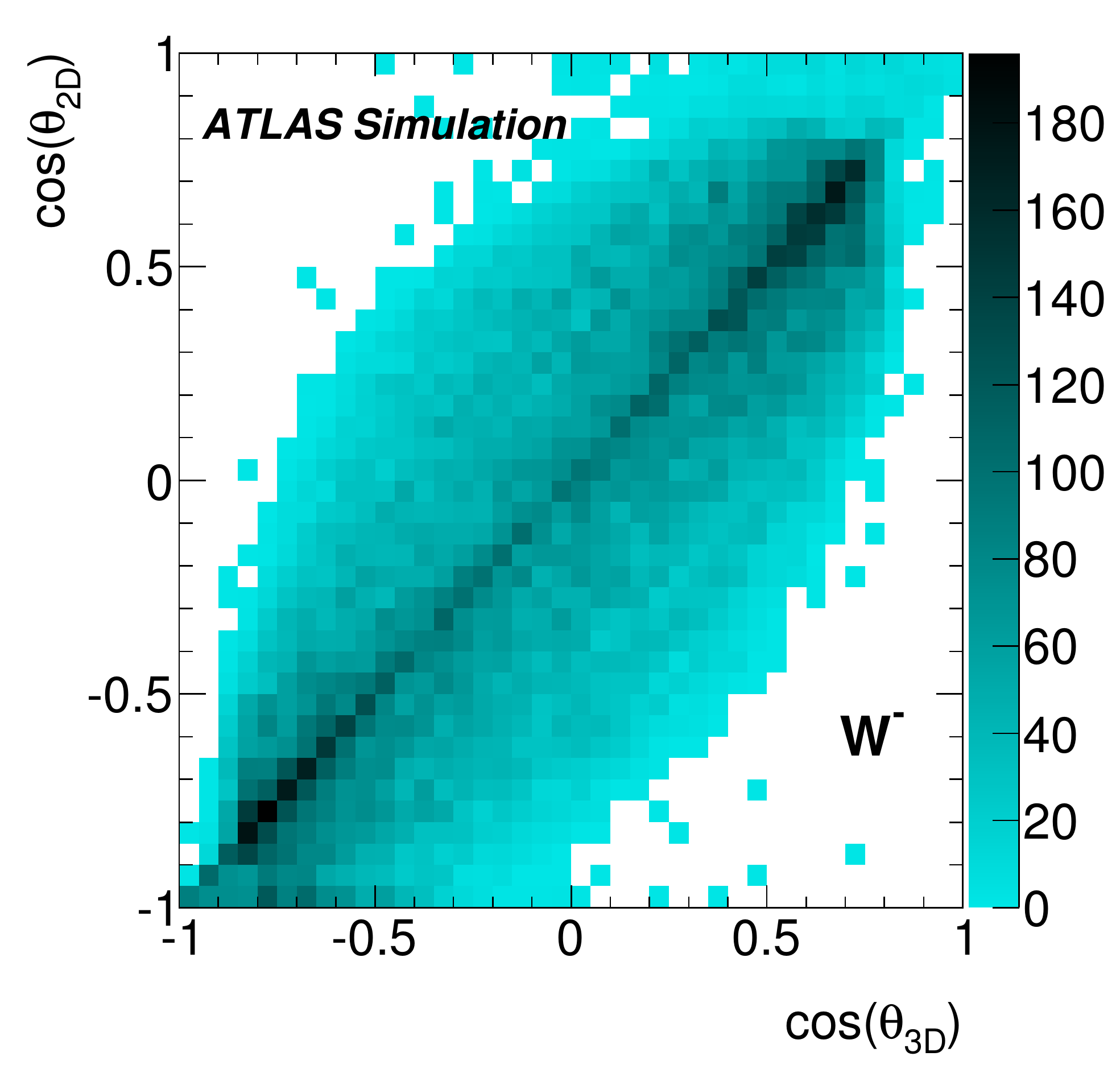}
      \label{fig:ctdres2}
    }
  \end{changemargin}
  \caption{Representation of \ctdd\ as a function of \ctd\ in events where the \Wboson\ transverse momentum is greater 
    than 50 \GeV, for (a) positive and (b) negative leptons. Events are simulated with \mcnlo\ after applying the acceptance
    and \trm\ cuts, as defined in Section\ \ref{sect:selection}.\label{fig:ctdresidu}}
\end{figure}

The measurement of helicity fractions is made by fitting \ctdd\ distributions with a weighted sum of templates obtained from Monte Carlo 
simulations, which correspond to longitudinal, left- and right-handed states. This is described in detail in Section\ \ref{sect:templatesandclosure}.

\section{Detector, data and simulation}
\label{sect:atlas}
\subsection{The ATLAS detector}

The ATLAS detector\ \cite{atlas} at the LHC covers nearly the entire solid angle around the collision region. It consists of an inner 
tracking system surrounded by a thin superconducting solenoid, electromagnetic and hadronic calorimeters, and a muon spectrometer 
incorporating three large superconducting toroid magnets.\par
The inner detector (ID) is immersed in a 2~T axial magnetic field and allows charged particle tracking in the range $|\eta| < 
2.5$. The high-granularity silicon pixel detector covers the vertex region and typically provides three measurements per track. It is 
followed by the silicon microstrip tracker which usually provides four two-dimensional measurement points per track. These silicon 
detectors are complemented by the transition radiation tracker, which enables radially extended track reconstruction up to $|\eta| = 2.0$. 
The transition radiation tracker also provides electron identification information based on the fraction of hits (typically 30 per track) 
above an energy threshold corresponding to transition radiation.\par
The calorimeter system covers the pseudorapidity range $|\eta|< 4.9$. Within the region $|\eta|< 3.2$, electromagnetic calorimetry is based on 
barrel and end-cap high-granularity lead liquid-argon (LAr) electromagnetic calorimeters, with an additional thin LAr presampler covering 
$|\eta| < 1.8$ to correct for energy loss in material upstream of the calorimeters. Hadronic calorimetry is provided by a steel/scintillating-tile
detector, segmented into three structures within $|\eta| < 1.7$, and two copper/LAr hadronic endcap calorimeters. The solid angle 
coverage is completed with forward copper/LAr and tungsten/LAr calorimeter modules optimised for electromagnetic and hadronic measurements 
respectively.\par
The muon spectrometer (MS) comprises separate trigger and high-precision tracking chambers measuring the deflection of muons in a magnetic 
field generated by superconducting air-core toroids. The precision chamber system covers the region $|\eta| < 2.7$, with three layers of 
monitored drift tubes complemented by cathode strip chambers in the region beyond $|\eta| = 2.0$ where the background is highest. The muon 
trigger system covers the range $|\eta| < 2.4$ with resistive plate chambers in the barrel, and thin gap chambers in the endcap regions.\par
A three-level trigger system is used to select interesting events\ \cite{Aad2011xs}. The Level-1 trigger is implemented in hardware and uses 
a subset of detector information to reduce the event rate to a design value of at most 75~kHz. This is followed by two software-based trigger 
levels which together reduce the event rate to about 200~Hz.

\subsection{Data sample}

\label{sect:datasample}
The data used in this analysis were collected from August to October 2010. Requirements on beam, detector and trigger conditions, as well as
on data quality, were used in the event selection, resulting in integrated luminosities of 37.3 \ipb\ for the electron channel and 31.4 \ipb\ 
for the muon channel (data where the muon trigger conditions varied too rapidly were not included). \par

\begin{table*}
  \centering
  \caption{Numbers of events in data and signal Monte Carlo samples, after standard and analysis cuts (see text), classified according to lepton flavour 
    and charge. The remaining numbers of events after standard plus analysis cuts are also represented as a percentage of the numbers of events 
    passing the standard selection.}
  \label{tab:Statistics}
  \footnotesize
  \begin{tabular}{|c|c|c|c||c|c|}
    \hline
    \multicolumn{2}{|c|}{ }& \mupl & \mumi & \epl & \emi \\
    \hline \hline   
    \multirow{3}{*}{Data} & Standard cuts & 79713 & 52186 & 67130 & 45690 \\
    &Analysis cuts (35 $<$ $\ptw$ $<$ 50 \GeV) & 4459 (5.6\%) & 3018 (5.8\%) &   3778 (5.6\%) & 2656 (5.8\%) \\
    &Analysis cuts ($\ptw$ $\ge$ 50 \GeV) & 3921 (4.9\%) & 2640 (5.1\%) &   3573 (5.3\%) & 2572 (5.6\%) \\
    \hline 
    \multirow{3}{*}{\mcnlo} & Standard cuts & 1484062 & 1041818 &   1054705 & 774952 \\
    &Analysis cuts (35 $<$ $\ptw$ $<$ 50 \GeV) & 76807 (5.2\%) & 52781 (5.1\%) &   54044 (5.1\%) & 39528 (5.1\%) \\
    &Analysis cuts ($\ptw$ $\ge$ 50 \GeV) & 57699 (3.9\%) & 39114 (3.8\%) &   43509 (4.1\%) & 31283 (4.0\%) \\
    \hline
    \multirow{3}{*}{\pow} & Standard cuts & 1498352 & 1056697 &   1056561 & 775894 \\
    &Analysis cuts (35 $<$ $\ptw$ $<$ 50 \GeV) & 82174 (5.5\%) & 59788 (5.7\%) &   58423 (5.5\%) & 44276 (5.7\%) \\
    &Analysis cuts ($\ptw$ $\ge$ 50 \GeV) & 66674 (4.5\%) & 47115 (4.6\%) &   50705 (4.8\%) & 37792 (4.9\%) \\
    \hline 
  \end{tabular}
\end{table*}
The integrated luminosity measurement has an uncertainty of 3.4\%\ \cite{Lumi,Aad2011dr}.\par

\subsection{Simulation}
\label{MonteCarlo}

Signal and background samples were processed through a \textsc{geant}4\ \cite{geant4} simulation of the ATLAS detector\ \cite{simulation} and 
reconstructed using the same analysis chain as the data.\par
The signal samples were generated using \mcnlo~3.4.2\ with \textsc{herwig}\ \cite{herwig} parton showering, and with \pow~1.0\ and \pyth\ parton
showering. Both used the CTEQ~6.6\ \cite{Pumplin2002vw} PDF set. All background samples were generated with \pyth~6.4.21\ \cite{pythia} 
except \ttbar\ for which \mcnlo\ was used. In order to study the sensitivity of the angular distributions to different NLO PDF sets,
the \mcnlo\ sample was reweighted\ \cite{Tricoli2005nx} according to MSTW~2008\ \cite{MSTW2008} and HERAPDF~1.0\ \cite{2009wt} PDF sets. \par
The radiation of photons from charged leptons was simulated using PHOTOS\ \cite{Photos}, and TAUOLA\ \cite{Tauola} was used for $\tau$ 
decays. The underlying event\ \cite{jimmy} was simulated according to the ATLAS tune\ \cite{mbtuneJan}. The Monte Carlo samples were generated with, 
in average, two soft inelastic collisions overlaid on the hard-scattering event. Events were subsequently reweighted so that the distribution
of the number of reconstructed vertices matched that in data, which was 2.2 on average.\par

\subsection{Event selection}
\label{sect:selection}
Events in this analysis are first selected using either a single-muon trigger with a requirement on the transverse momentum \ptl\ of at least 
13 \GeV, or a single-electron trigger, with a \ptl\ requirement of at least 15 \GeV\ \cite{Aad2011xs}. Subsequent selection criteria closely 
follow those used for the \Wboson\ boson inclusive cross-section measurement reported in Ref.\ \cite{Aad2011dm}.\par
Events from $pp$ collisions are selected by requiring a reconstructed vertex compatible with the beam-spot position and with at least three 
associated tracks each with transverse momentum greater than 0.5 \GeV.\par
Electron candidates are required to satisfy \ptl\ $>$ 20 \GeV, $|\eta|$ $<$ 2.47 (but removing the region where barrel and endcap calorimeters 
overlap, i.e. 1.37 $<$ $|\eta|$ $<$ 1.52) and to pass the ``tight'' identification criteria described in Ref.\ \cite{Aad2011mk}. This selection rejects 
charged hadrons and secondary electrons from conversions by fully exploiting the electron identification potential of the detector. It makes 
requirements on shower shapes in the electromagnetic calorimeter, on the angular matching between the calorimeter energy cluster and the ID
track, on the ratio of cluster energy to track momentum, and on the number of hits in the pixels (in particular a hit in the innermost layer 
is required), in the silicon microstrip tracker and in the transition radiation tracker. \par
Muon candidates are required to be reconstructed in both the ID and the MS, with transverse momenta satisfying the conditions $|(p_{\rm T}^{\rm MS}-
p_{\rm T}^{\rm ID})/p_{\rm T}^{\rm ID}|<0.5$ and $p_{\rm T}^{\rm MS}>10\ \GeV$. The two measurements are then combined, weighted by their respective 
uncertainties, to form a \emph{combined muon}. The \Wboson\ candidate events are required to have at least one combined muon track with \ptl\ $>$ 
20 \GeV, within the range  $|\eta|$ $<$ 2.4. This muon candidate must also satisfy the isolation condition $(\Sigma p_{\rm T}^{\rm ID})/\ptl$ $<$ 0.2, 
where the sum is over all charged particle tracks around the muon direction within a cone of size $\Delta R$ = $\sqrt{(\Delta\eta)^2 + (\Delta\phi)^2}$ 
= 0.4. Finally, to reduce the contribution of cosmic-ray events, and beam-halo induced by proton losses from the beam, the analysis requires the 
reconstructed vertex position along the beam axis to be within 20 cm of the nominal interaction point.\par
The \MTE\ (\met) is reconstructed as the negative vector sum of calibrated ``objects'' (jets, electrons or photons, muons) to which the energies 
of calorimeter cells not associated to any of the objects are added. \met\ is required to be larger than 25 \GeV. A cut \trm\ $>$ 40 \GeV\ is 
finally applied.\par
In addition to these cuts, called in the following \emph{standard cuts}, additional selections are used for this analysis. A low \trm\ cut at 50\ 
\GeV\ is applied to minimise backgrounds, and a high \trm\ cut at 110 \GeV\ is applied to remove tails of badly reconstructed events. Finally a 
\ptw\ selection in two bins (35 $<$ $\ptw$ $<$ 50 \GeV, and \ptw $>$ 50 \GeV) is made. The numbers of events passing these cuts are shown in 
\Tab{Statistics}.\par

\begin{table*}
  \caption{Background fractions (with respect to the expected signal) obtained from Monte Carlo simulations (electroweak and \ttbar) normalised 
    to state-of-the-art signal cross-section predictions (see text) and from data (jet background) by fitting \met\ distributions with templates. 
    \label{tab:QCDfractions}}
  \centering  
  \footnotesize
  \begin{tabular}{|cc|c|c||c|c|}
    \cline{2-6}
    \multicolumn{1}{c|}{}&Fractions (\%) & $\mu^+$ & $\mu^-$ & e$^+$ & e$^-$ \\ 
    \hline
    \multirow{5}*{Standard cuts}     & jet & 2.1 $\pm$ 0.1 & 3.1 $\pm$ 0.2 & 2.4 $\pm$ 0.1 & 3.6 $\pm$ 0.1 \\
    & \ttbar  & 0.2 & 0.4 & 0.3 & 0.5\\
    & \Wtaun  & 2.6 & 2.8 & 2.3 & 2.5\\
    & \Ztautau  & 0.1 & 0.2 & 0.1 & 0.1\\
    & \Zll  & 2.9 & 3.9 & 0.1 & 0.2\\
    \hline    
    \multirow{5}*{\minitab[c]{Analysis cuts\\ (35 $<$ \ptw\ $<$ 50 \GeV)}} & jet & 2 $\pm$ 2 & 2 $\pm$ 2 &  2.4 $\pm$ 0.4 & 2.5 $\pm$ 0.5 \\
    & \ttbar  & 0.5 & 0.7 & 0.6 & 0.9 \\
    & \Wtaun  & 2.1 & 2.4 & 1.8 & 1.9 \\
    & \Ztautau  & 0.1 & 0.1 & 0.1 & 0.1\\
    & \Zll  & 2.9 & 3.9 & 0.3 & 0.4\\
    \hline
    \multirow{5}*{\minitab[c]{Analysis cuts \\(\ptw\ $>$ 50 \GeV)}} & jet & 2 $\pm$ 2 & 2 $\pm$ 2 &  2.0 $\pm$ 0.3 & 2.5 $\pm$ 0.4 \\
    & \ttbar  & 2.8 & 4.1 & 3.5 & 5.0\\
    & \Wtaun  & 2.1 & 2.0 & 1.9 & 2.0 \\
    & \Ztautau  & 0.1 & 0.1 & 0.1 & 0.1 \\
    & \Zll  & 2.6 & 3.5 & 0.3 & 0.4 \\
    \hline
  \end{tabular}
\end{table*}

The data are compared to expectations based on Monte Carlo simulations. In addition to the signal (\Wboson\ production followed by leptonic 
decay to an electron or a muon), the following electroweak backgrounds are considered: \Wtaun, \Zee, \Zmm\ and \Ztautau, as well as \ttbar\ 
events with at least one semi-leptonic decay. Jet production via QCD was also simulated, but the final estimate of this background is obtained 
from data, as explained in Section\ \ref{sect:backgrounds}.\par

\section{Signal normalisation and background estimate}
\label{sect:normalisatiom}
\subsection{Signal normalisation}

The \Wln\ production cross-sections and the decay branching ratios used in this study are normalised to the NNLO predictions of the FEWZ program\ 
\cite{fewz} with the MSTW~2008 PDF set:
\begin{linenomath*}
  \begin{gather*}
    \hspace*{2.8cm} \sigma^{NNLO}_{\ensuremath{W^+} \rightarrow \ell \nu} = 6.16\ \mathrm{nb}\,, \hfill \\
    \hspace*{2.8cm} \sigma^{NNLO}_{\ensuremath{W^-} \rightarrow \ell \nu} = 4.30\ \mathrm{nb}\,. \hfill     
  \end{gather*}
\end{linenomath*}
The estimated uncertainties on each cross-section coming from the factorisation and renormalisation scales as well as from the parton distribution 
functions are expected to be approximately 5\%\ \cite{Aad2011dm}.

\subsection{Background estimates}
\label{sect:backgrounds}
\Wboson\ events decaying into $\tau$-leptons with subsequent leptonic $\tau$ decays contribute as background to both electron and muon channels. 
Contributions from \Zmm\ decays are significant in the muon channel, where the limited $\eta$ coverage of the tracking and muon systems can result 
in fake \met\ when one of the muons is missed. On the contrary, the \Zee\ background is almost negligible in the electron channel due to the nearly 
hermetic calorimeter coverage over $|\eta|< 4.9$. For both the electron and the muon channels, contributions from \Ztautau\ decays and from 
\ttbar\ events involving at least one leptonic \Wboson\ decay are also taken into account. The latter is particularly relevant for the large 
transverse momentum \Wbosons\ studied here.\par 
The normalisation of electroweak and \ttbar\ backgrounds is based on their total theoretical cross-sections. These cross-sections are calculated 
at NLO (plus next-to-next-to-leading-log corrections) for \ttbar\ \cite{topxs,Langenfeld2009tc}, and at NNLO for the others. The contributions of 
these backgrounds to the final data sample have been estimated using simulation to model acceptance effects.\par
One of the major background contributions, especially in the electron channel, is from dijet production via QCD processes. The selected leptons 
from these processes have components from semi-leptonic decays of heavy quarks, hadrons mis-identified as leptons, and, in the case of the 
electron channel, electrons from conversions. The \MTE\ is due mainly to jet mis-measurement. For both the electron and muon channels, these 
sources of background are obtained from the data. Monte Carlo simulated samples are also used for cross-checks.\par

The jet background is obtained by fitting the \met\ data distributions to the sum of the \Wln\ signal and the electroweak and \ttbar\ backgrounds, 
normalised as described above and called hereafter the ``electroweak template'', plus a ``jet event template'' derived from control samples in the 
data.\par
In the electron case, the jet event template is obtained by selecting electron candidates passing the ``loose'' selection\ \cite{Aad2011mk}, but 
failing one or more of the additional criteria required to flag an electron as ``medium'' as well as an isolation cut (which removes signal 
events).\par
In the muon case, the jet event template is obtained by inverting the track isolation requirement.\par
In both cases, the relative normalisation of the jet event and electroweak templates is determined by fitting the two templates to the 
\met\ distribution in the data down to 10 \GeV. The jet event fraction is then obtained from the (normalised) jet event template
by counting events above \met=\ 25 \GeV.\par
The background fractions determined with the methods described above, for the standard cuts and for the standard plus analysis cuts, 
are shown in \Tab{QCDfractions}. These results were obtained with \mcnlo\ for the signal simulation, and are in agreement with those obtained with 
\pow. For the muon channel, as jet event fractions are small and measured with larger uncertainties than for electrons, a value of 2\% with an 
uncertainty of $\pm$ 2\% is used for both \Wp\ and \Wm. \Tab{QCDfractions} shows the statistical uncertainties from the jet template method. 
Uncertainties on the measurement due to background modelling are described in Section\ \ref{sect:backnorm}.\par

\section{Data to Monte Carlo comparison of transverse helicity}
\label{sect:datamccomp}

\begin{figure*}
  \begin{changemargin}{-0.05\textwidth}{-0.05\textwidth}
    \centering
    \subfigure[$\mu^+$]{
      \includegraphics[width=0.38\textwidth]{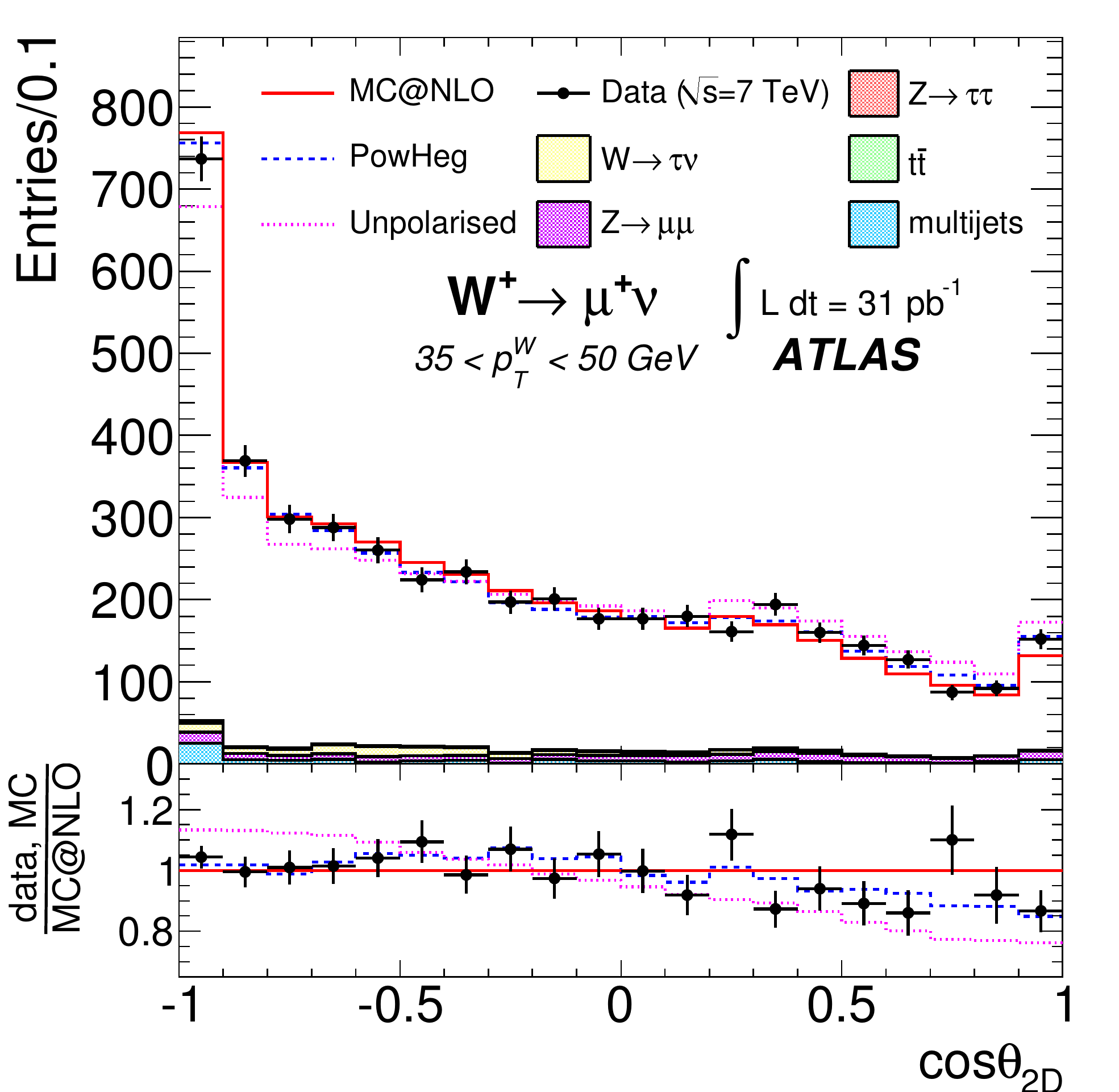}
    }\hspace*{-0.5cm}
    \subfigure[$\mu^-$]{
      \includegraphics[width=0.38\textwidth]{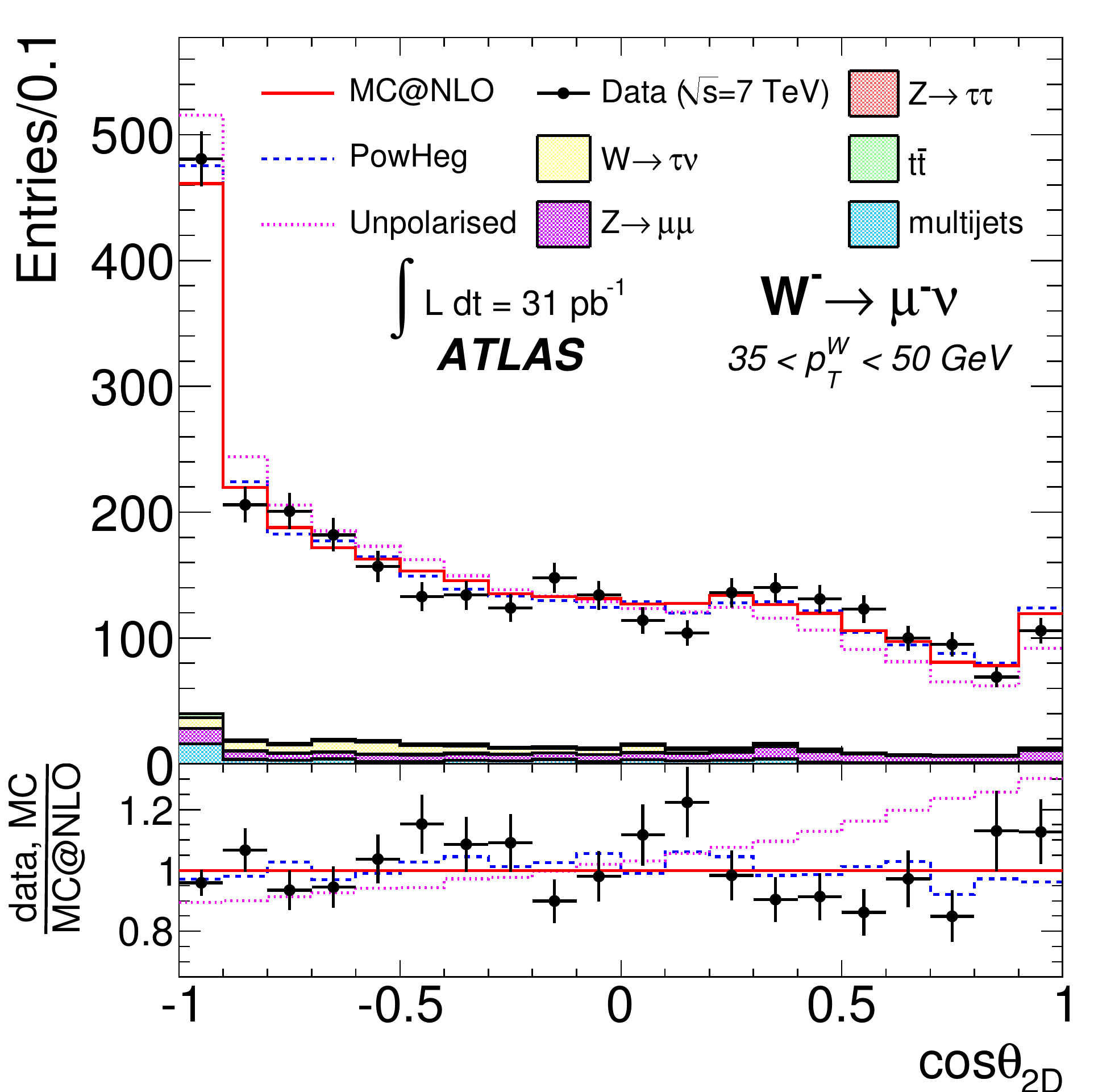}
    }
    \subfigure[e$^+$]{
      \includegraphics[width=0.38\textwidth]{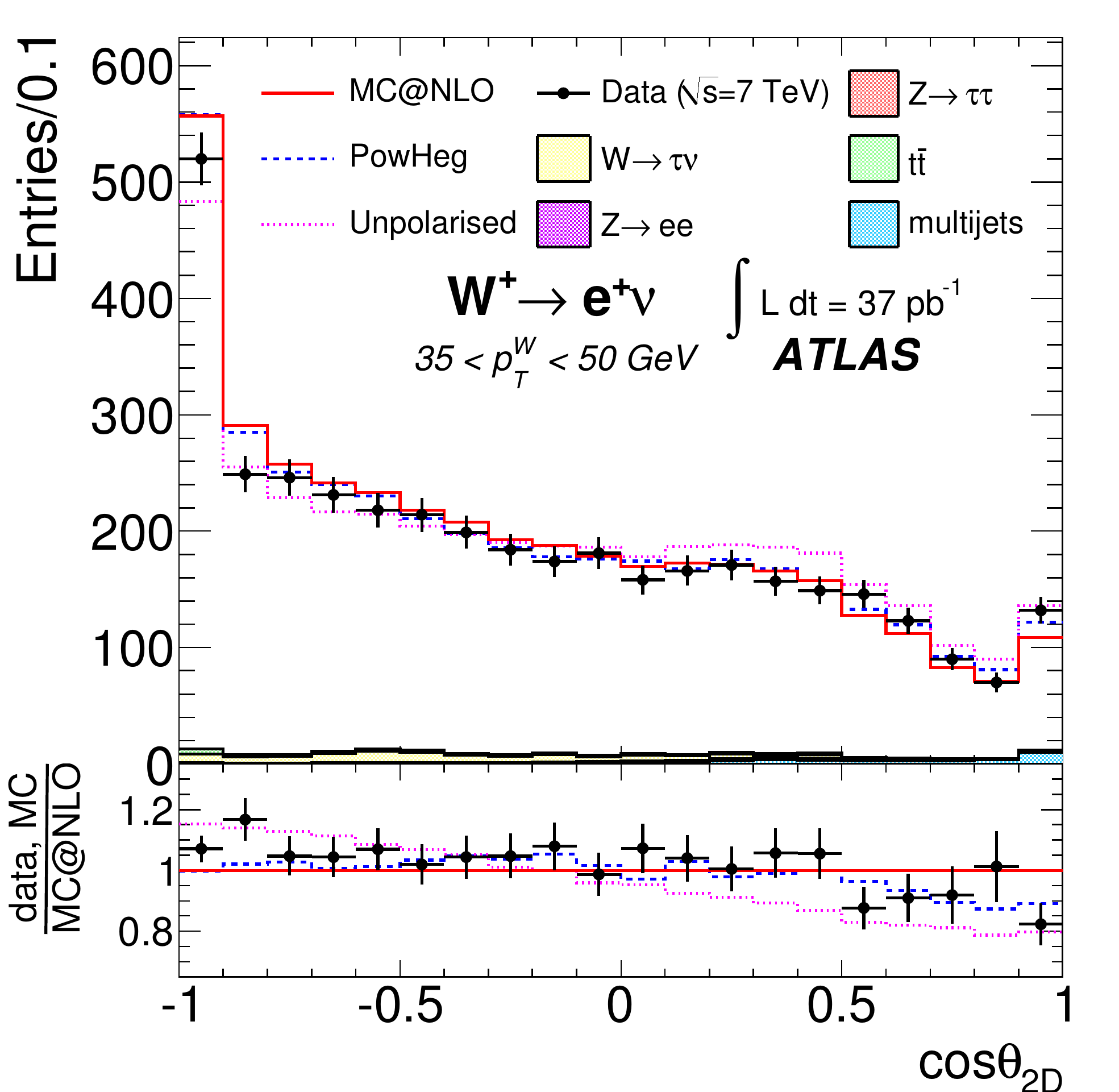}
    }\hspace*{-0.5cm}
    \subfigure[e$^-$]{
      \includegraphics[width=0.38\textwidth]{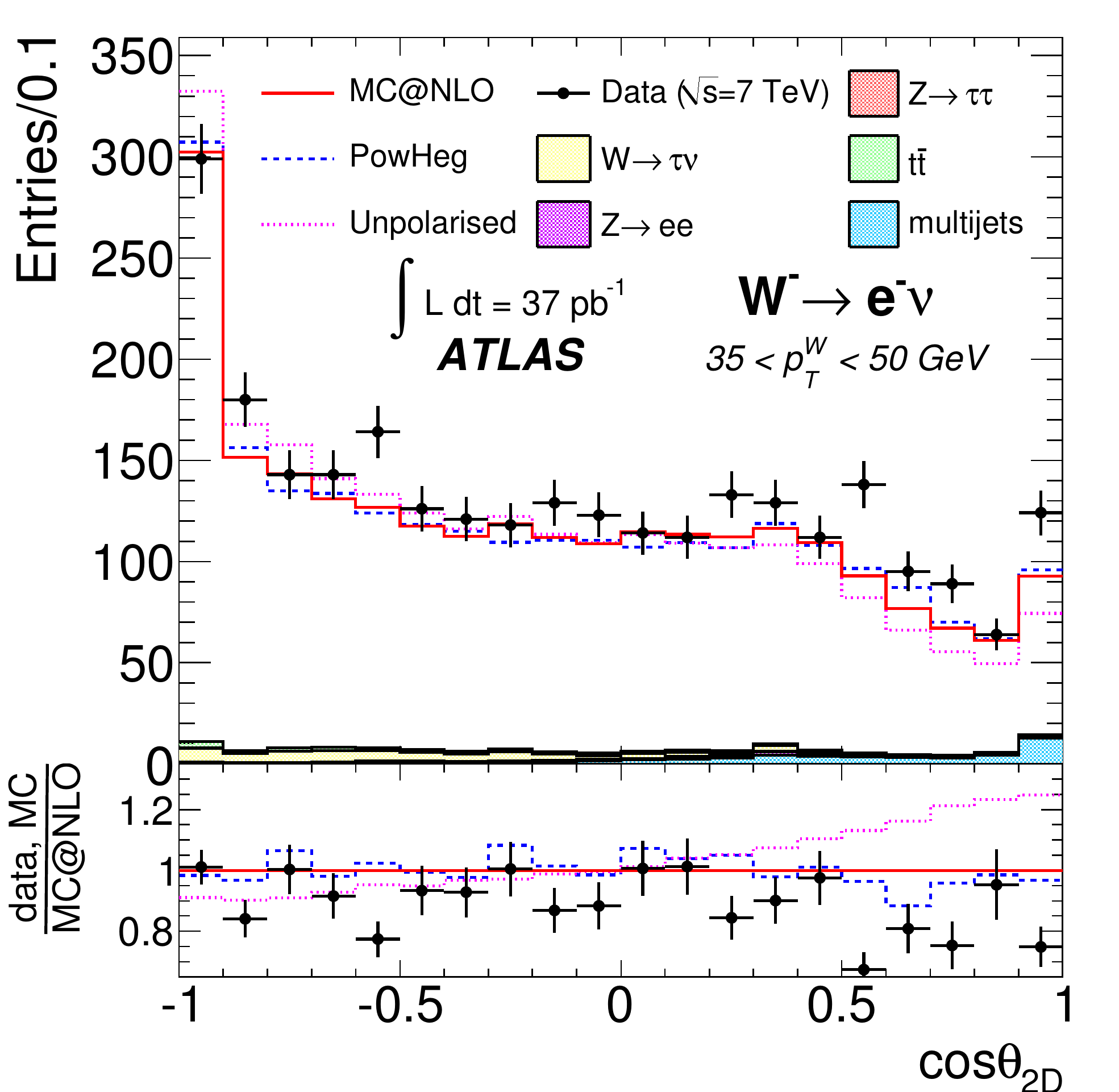}
    }
  \end{changemargin}
  \caption{The \ctdd\ distributions for 35 $<$ \ptw\ $<$ 50 \GeV. The data (dots) are compared to the distributions from \pow\ (dashed line), 
    \mcnlo\ (solid line), and for unpolarised \Wbosons\ (dotted line) in the muon (top) and electron (bottom) channel, 
    split by charge. The bottom parts of each plot represent the ratio of data, \pow\ and unpolarised distributions to \mcnlo.
    \label{fig:NoTemplFitEl35}}
\end{figure*}

As shown in Ref.\ \cite{Aad2011dm}, \mcnlo\ and \pow\ give a rather good description of inclusive \Wboson\ production. However both generators 
were shown\ \cite{Belloni1361975} to underestimate the fraction of events at large \ptw\ (see also \Tab{Statistics}). While this affects the 
relative fraction of data versus Monte Carlo events retained in the two \ptw\ bins of the analysis, it should not significantly impact the angular 
distributions used to measure the \Wboson\ polarisation. This is discussed in more detail in Section\ \ref{sect:ptwrew}. 

\begin{figure*}
  \begin{changemargin}{-0.05\textwidth}{-0.05\textwidth}
    \centering
    \subfigure[$\mu^+$]{
      \includegraphics[width=0.38\textwidth]{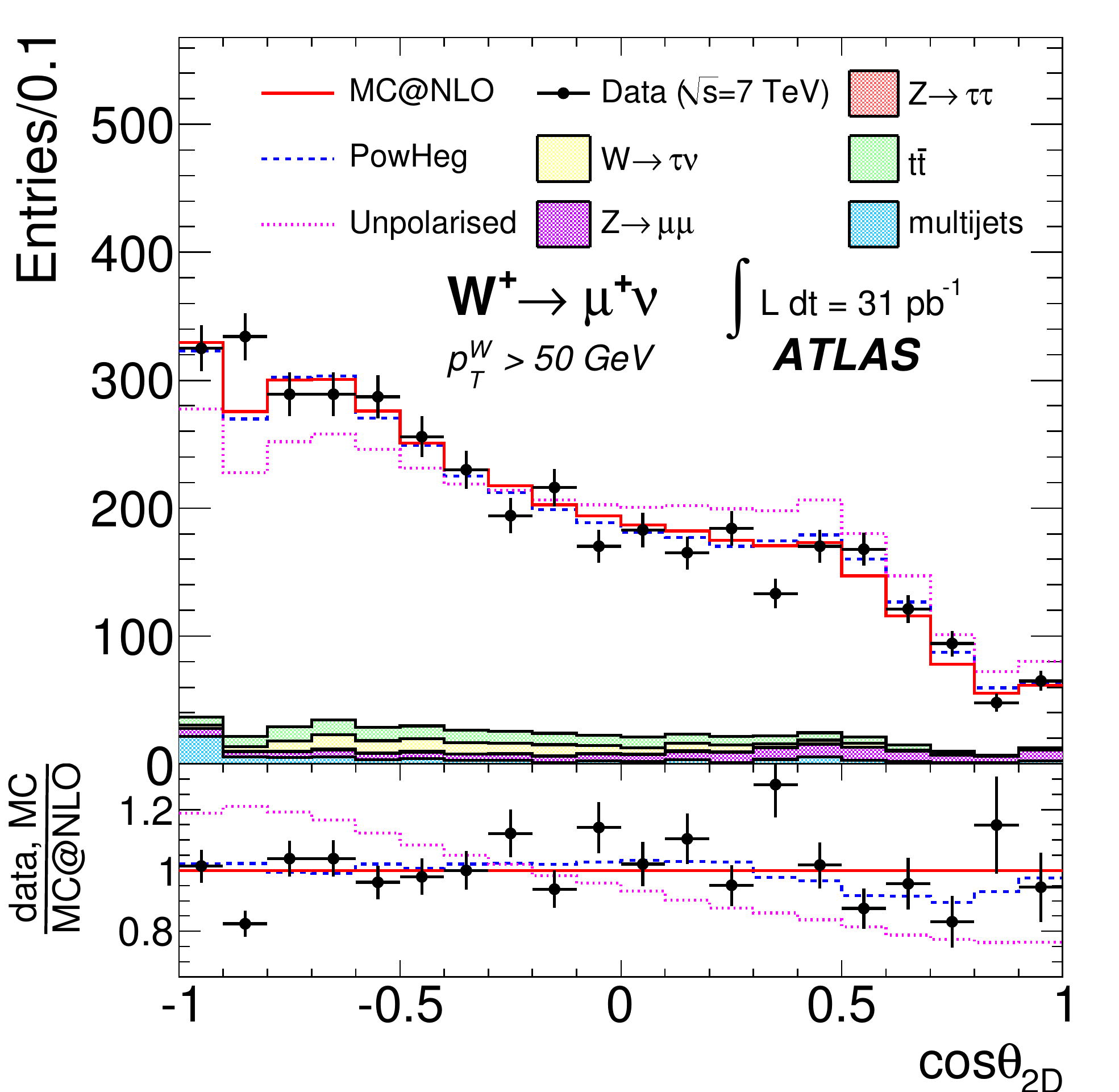}
    }\hspace*{-0.5cm}
    \subfigure[$\mu^-$]{
      \includegraphics[width=0.38\textwidth]{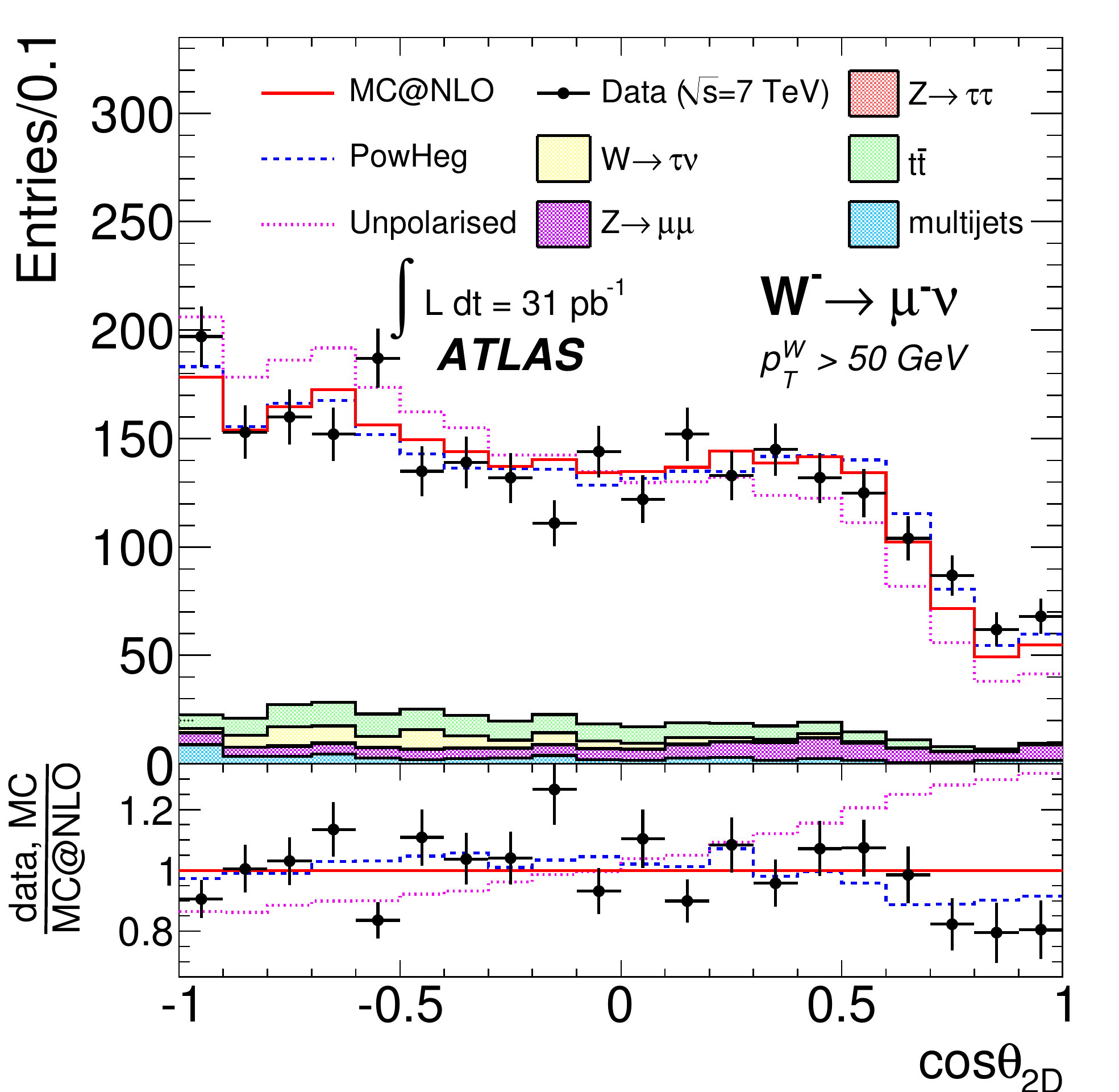}
    }
    \subfigure[e$^+$]{
      \includegraphics[width=0.38\textwidth]{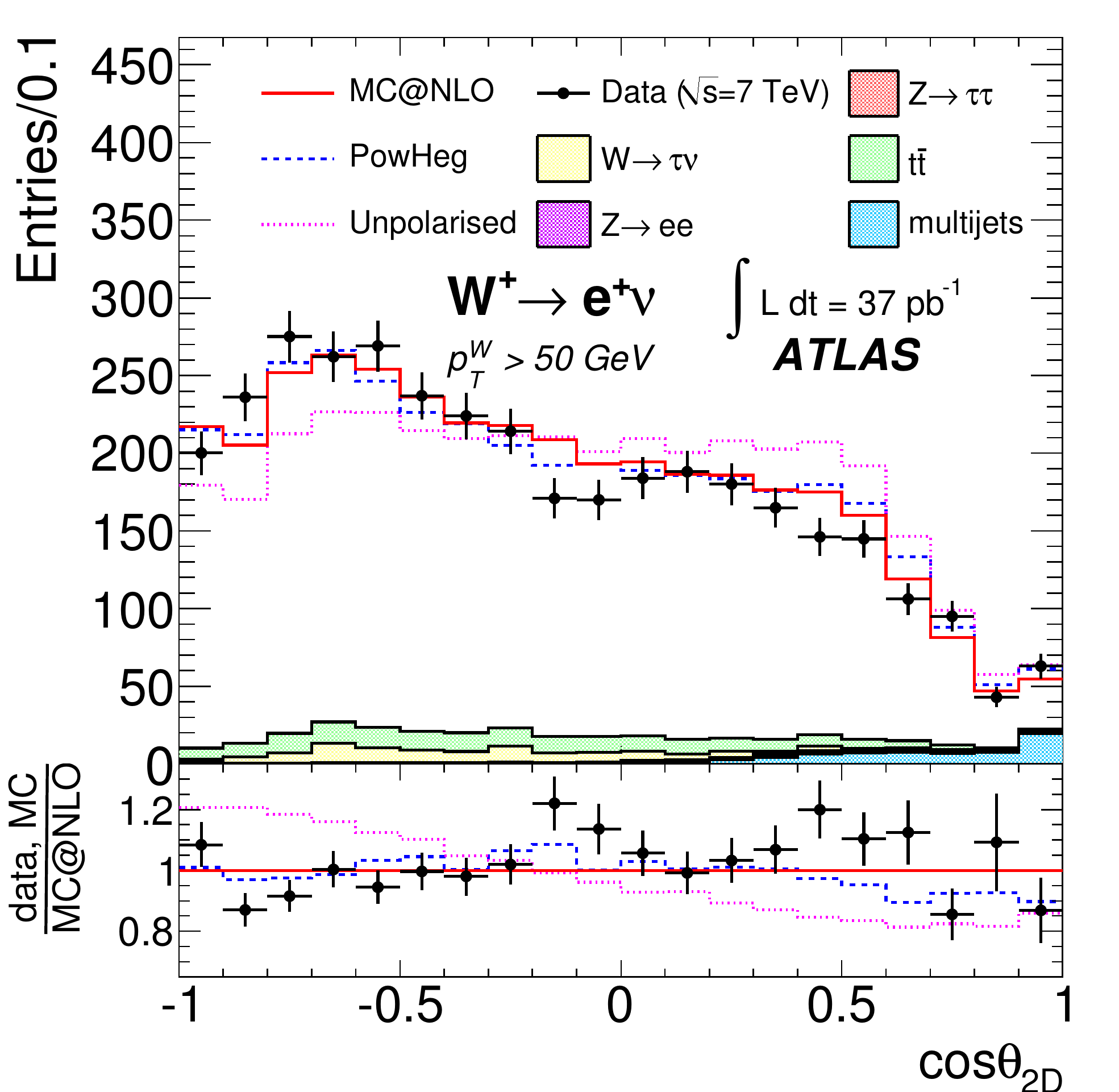}
    }\hspace*{-0.5cm}
    \subfigure[e$^-$]{
      \includegraphics[width=0.38\textwidth]{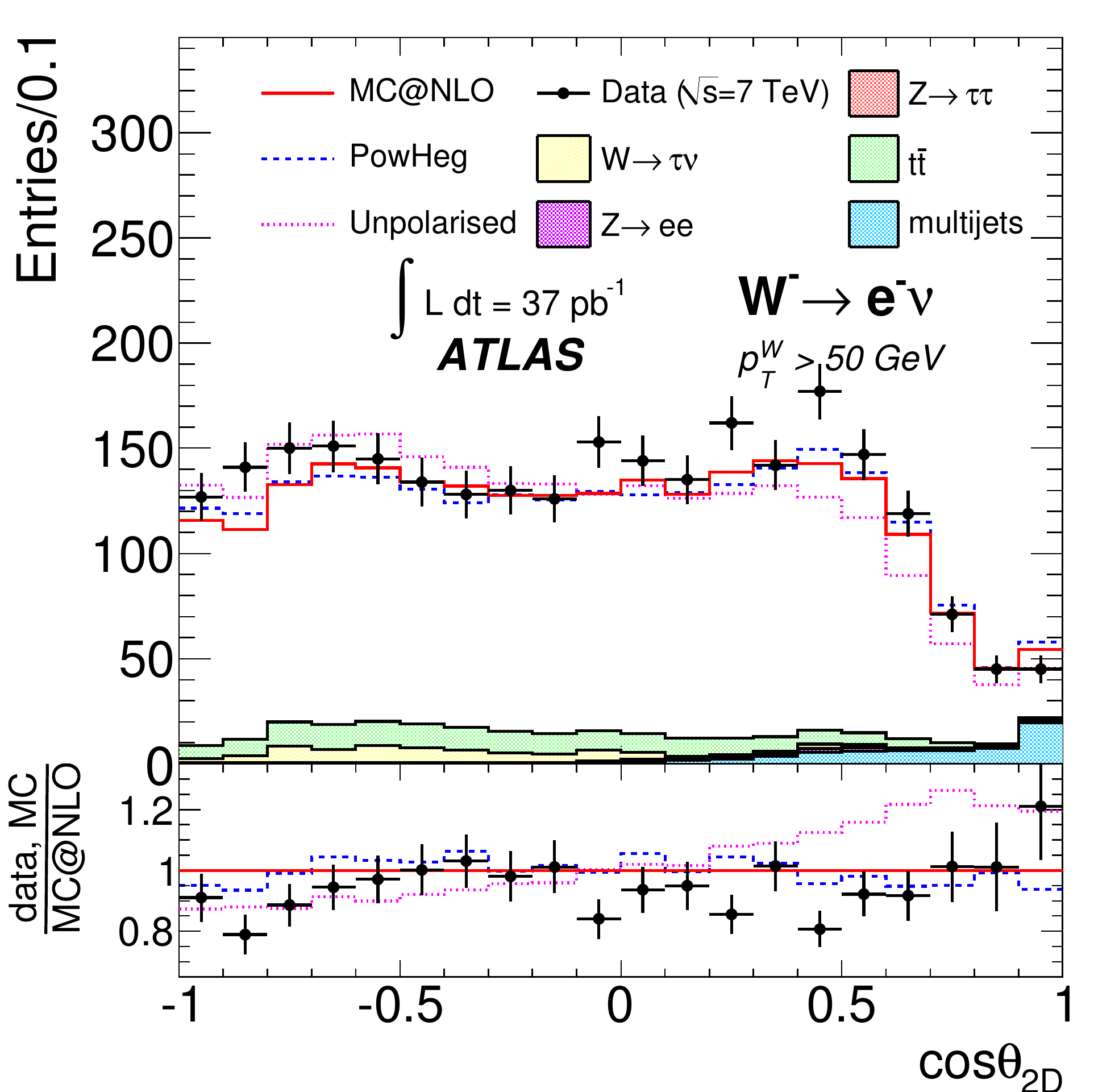}
    }
  \end{changemargin}
  \caption{The \ctdd\ distributions for \ptw\ $>$ 50 \GeV. The data (dots) are compared to the distributions from \pow\ (dashed line), 
    \mcnlo\ (solid line), and for unpolarised \Wbosons\ (dotted line) in the muon (top) and electron (bottom) channel, split 
    by charge. The bottom parts of each plot represent the ratio of data, \pow\ and unpolarised distributions to \mcnlo. \label{fig:NoTemplFitEl}}
\end{figure*}
\begin{table*}
\centering
  \caption{The $\chi^2$ values from the comparison of the data with the \mcnlo, \pow\  and unpolarised predictions for the \ctdd\ distributions 
    (see \Figs[NoTemplFitEl35]{NoTemplFitEl}). The number of degrees of freedom in the fits is 19. Only statistical uncertainties are considered.}
  \begin{tabular}{|c|c|c|c|c||c|c|c|c|}
    \hline
    \multicolumn{1}{|c|}{\multirow{2}{*}{\minitab[c]{$\chi^2$ between\\ data and}}}& \multicolumn{4}{c||}{35 $<$ \ptw\ $<$ 50 \GeV \bigstrut} & \multicolumn{4}{c|}{\ptw\ $>$ 50 \GeV} \\
    \cline{2-9}
    & $\mu^+$ & $\mu^-$ & $e^+$ & $e^-$ & $\mu^+$ & $\mu^-$ & $e^+$ & $e^-$ \\
    \hline
    \mcnlo\  Monte Carlo & 20.0 & 25.0 & 17.0 & 32.1 & 36.2 & 31.5 & 28.6 & 17.3  \\ 
    \hline 
    \pow\  Monte Carlo & 12.8 & 22.9 & 10.7 & 25.5 & 40.3 & 32.7 & 30.3 & 16.3 \\ 
    \hline  
    Unpolarised  & 23.6 & 33.5 & 28.0 & 79.5 & 62.4 & 44.2 & 129.2 & 42.9 \\ 
    \hline 
  \end{tabular}
  \label{tab:chi2values}
\end{table*}

\begin{figure*}
  \begin{changemargin}{-0.05\textwidth}{-0.05\textwidth}
    \captionsetup{type=figure}
    \centering
    \subfigure[\fl\ for $\ell^+$]{
      \tikz{ \node (f0p) { \includegraphics[width=0.38\textwidth]{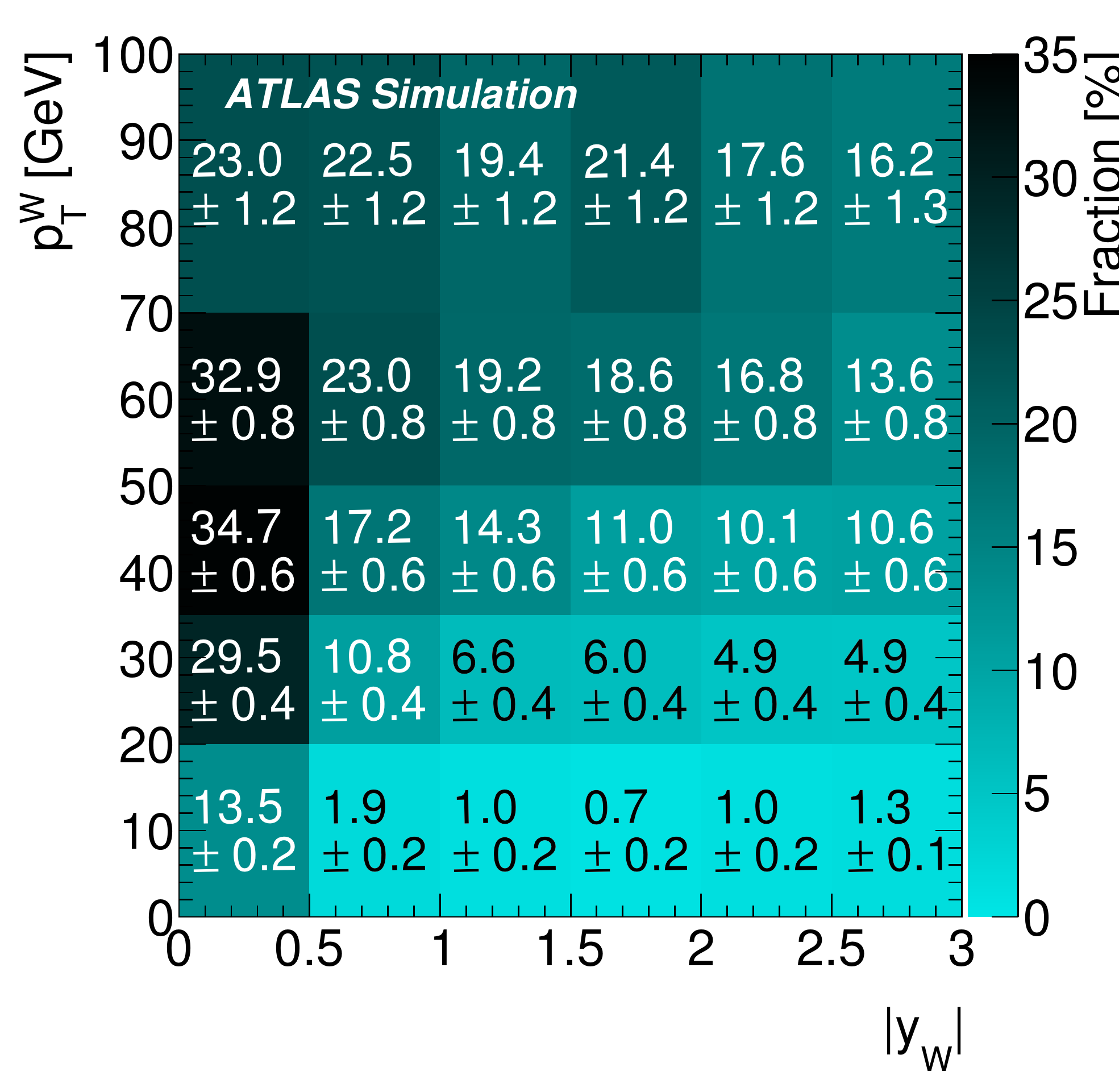}};}
    }
    \subfigure[\fl\ for $\ell^-$]{
      \tikz{ \node (f0m) { \includegraphics[width=0.38\textwidth]{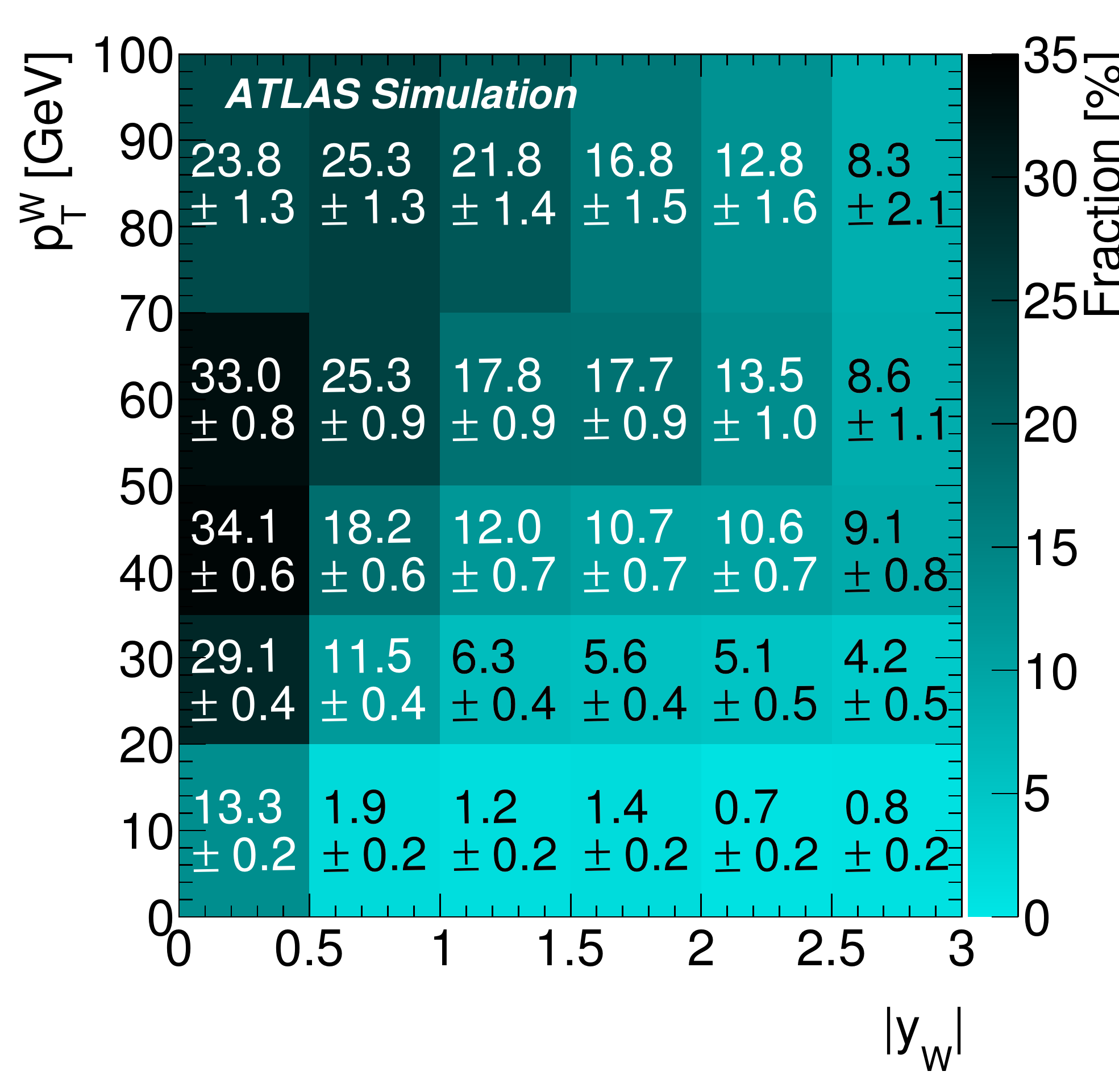}};}
    }

    \subfigure[\fLmR\ for $\ell^+$]{
      \tikz{ \node (fLmRp) { \includegraphics[width=0.38\textwidth]{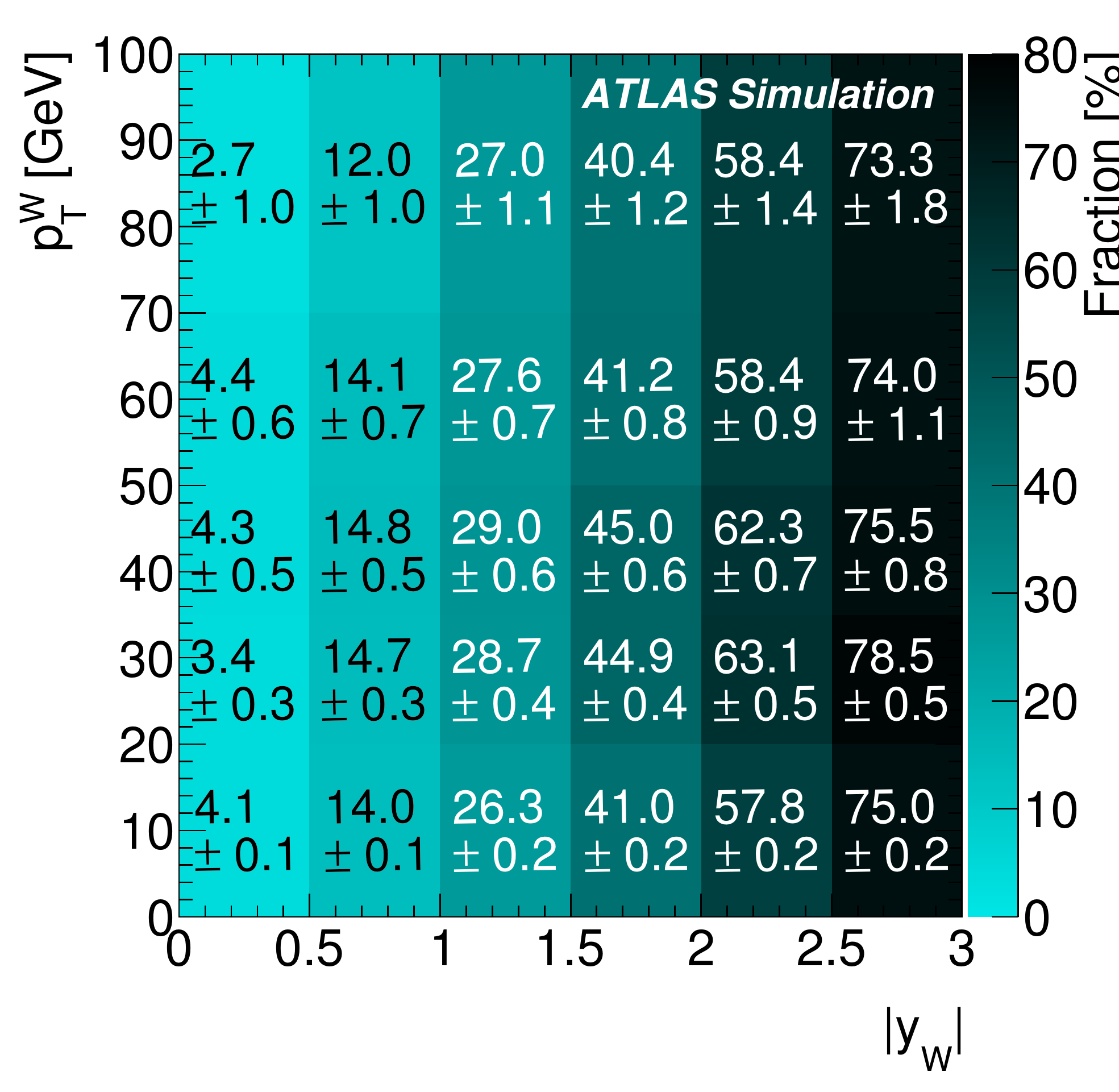}};}
    }
    \subfigure[\fLmR\ for $\ell^-$]{
      \tikz{ \node (fLmRm) { \includegraphics[width=0.38\textwidth]{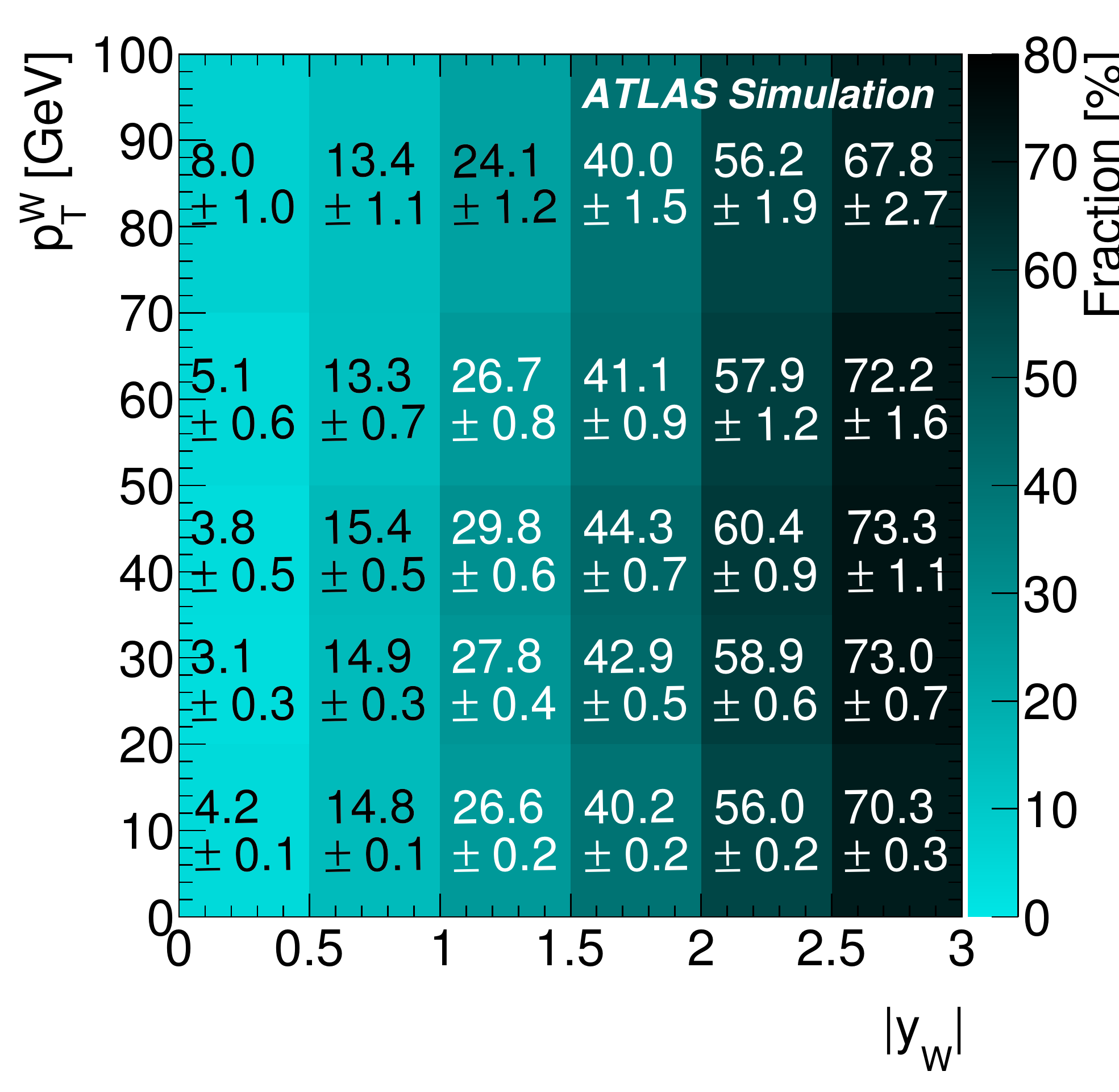}};}
    }
  \end{changemargin}
  \vspace*{0.3cm}
  \captionsetup{type=figure}
  \caption{Computed values of \fl\ (top) and \fLmR\ (bottom) using fits with \Eqn{fracdef} to \mcnlo\ samples in (\absyw, \ptw) bins, 
    split by charge. These values are used to calculate the weights needed to create helicity templates.\label{fig:ComputedFractions}}
  \centering 
\end{figure*}

\FFigs[NoTemplFitEl35]{NoTemplFitEl} show the \ctdd\ distributions for electrons and muons and both charges, compared to the predictions from 
\mcnlo\ and \pow\ and to the expected behaviour of unpolarised \Wbosons\ (the unpolarised distributions are obtained by averaging the longitudinal, 
left- and right-handed \mcnlo\ templates with equal weights, see Section\ \ref{templates}). The good agreement of both the \mcnlo\ and \pow\  
distributions with data is demonstrated also by the $\chi^2$ values reported in \Tab{chi2values}. It is also clear from \Tab{chi2values} and 
\Figs[NoTemplFitEl35]{NoTemplFitEl} that the production of unpolarised \Wbosons\ does not match the data.\par
For the electron channel, the jet background clusters around \ctdd=1, which supports the assumption that these were two-jet events, where one of 
the jets was mis-identified as an electron. On the other hand, in the muon channel, the jet background clusters around \mbox{\ctdd=$-$1}, in 
agreement with the assumption that the background originates mainly from semi-leptonic decay of heavy-flavour in jets. \par

\section{Helicity templates and Monte Carlo closure test}
\label{sect:templatesandclosure}

\subsection{Construction of helicity templates}
\label{templates}

In order to measure the helicity fractions, it is necessary to construct \ctdd\ distributions corresponding to samples of longitudinal, left- 
and right-handed \Wbosons\ that decay into a lepton and a neutrino. As a check at the generator-level, and for the correction procedure (see 
Section\ \ref{sect:systemp}), \ctd\ distributions corresponding to the three polarisation states were also made. All these distributions are 
called helicity templates in the following. The templates were built independently from \mcnlo\ and from \pow\  using the following reweighting 
technique.\par
It was first verified that, at the generator-level, and in bins of limited size in \ptw\ and \yw, \Wboson\ decays generated with the Monte Carlo
simulations are well described by \Eqn{fracdef}. The generator-level \ctd\ distributions were then fitted with the distribution corresponding to 
this equation, which gave the values of \fL, \fl\ and \fR\ in \yw\ and \ptw\ bins. The results, in terms of \fl\ and \fLmR, are shown in 
\Fig{ComputedFractions} for \mcnlo. The size of the bins results from a compromise between the rate of variation of the coefficients and the size 
of the available samples.\par
Several conclusions may be drawn from \Fig{ComputedFractions}. The longitudinal fraction, which is very small for low \ptw, grows with \ptw\ 
(especially at low \absyw), before flattening out and then starting to decrease. The difference between the fractions of left- and right-handed 
\Wboson\ bosons is small for low \absyw\ and grows quickly with \absyw, reaching up to 70\% for \absyw\ = 3. As already explained in Section\ 
\ref{sect:intro}, a smaller left-right difference is expected for negative than for positive \Wbosons; however in the \ptw\ 
range analysed here, these differences differ by at most a few percent. The analysis of systematic uncertainties described in Section\ 
\ref{sect:escale}, shows that it is experimentally advantageous to average the measured values of \fLmR\ between the two charges. As an 
anticipation of this observation, it can be seen in \Fig{ComputedFractions} that this averaging is physically meaningful.\par
\begin{figure}[h!]
  \begin{changemargin}{-0.05\textwidth}{0cm}
    \begin{center}
      \includegraphics[width=0.49\textwidth]{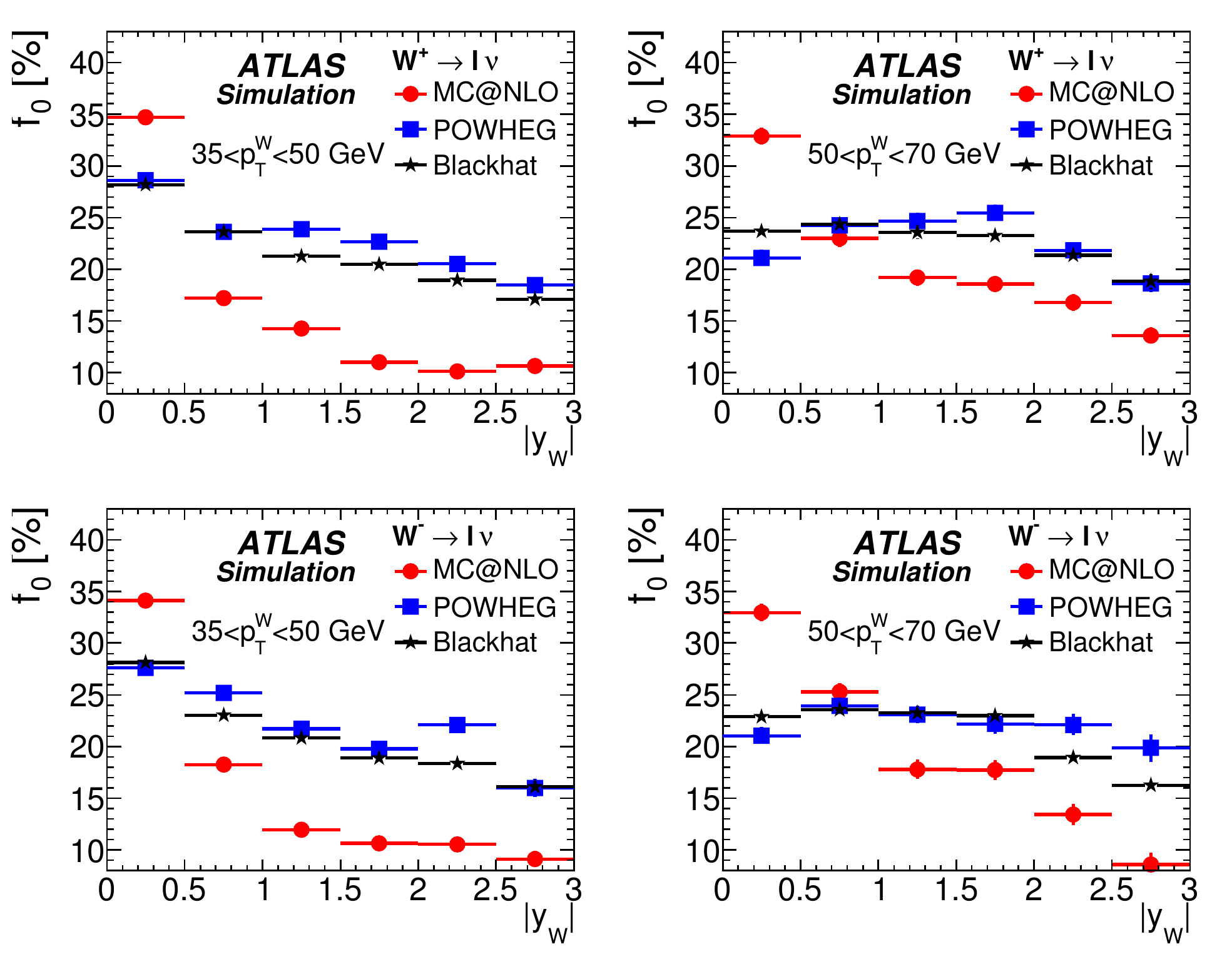}
    \end{center}
  \end{changemargin}
  \caption{Evolution of the longitudinal polarisation fraction as a function of \absyw, in \mcnlo, \pow\ and a calculation based on BlackHat, for
    \Wp\ (top) and \Wm\ (bottom) for two \ptw\ bins.\label{fig:CompSherpa}}
\end{figure}

An equivalent analysis for \pow\ shows a similar trend for \fLmR\ as observed for \mcnlo. For \fl, in the \ptw\ range analysed here, \pow\ 
exhibits a much flatter dependence on \yw\ than \mcnlo,  the average values being, however, very close to each other. Analytical calculations 
at NNLO reported in Ref.\ \cite{Bern2011ie} by the BlackHat collaboration are very close to \pow. This is illustrated in \Fig{CompSherpa}.\par

Samples representing longitudinal, left- and right-handed states are obtained by reweighting the \mcnlo\ or \pow\ simulated events according to :
\begin{changemargin}{-0.015\textwidth}{-0.015\textwidth}
  \begin{equation}
    \label{eqn:weicompu}
    \frac{ \left. \displaystyle\frac{1}{\sigma^\pm}\frac{d\sigma^\pm}{d\ctd} \right|_{L/0/R}}{\frac{3}{8}\fL (1\mp\ctd)^2 + \frac{3}{8}\fR (1\pm\ctd)^2 + \frac{3}{4}\fl \sct_{\rm 3D}}
  \end{equation}
\end{changemargin}
where 
\begin{equation}
\hfill  \left. \frac{1}{\sigma^\pm}\frac{d\sigma^\pm}{d\ctd} \right|
  \begin{array}{l}
    {\scriptstyle L}\\
    {\scriptstyle 0}\\
    {\scriptstyle R} 
  \end{array}
  = \frac{3}{8} \left \lbrace 
    \begin{array}{l} 
      (1\mp\ctd)^2 \\
      2 \std^2 \\
      (1\pm\ctd)^2 
    \end{array}
  \right . \hfill
\end{equation}
and where the denominator corresponds to the general form of the differential cross-section in which the coefficients are taken from 
\Fig{ComputedFractions} (or its equivalent from \pow), for the corresponding value of \ptw\ and \absyw. In these equations, the upper (lower) 
sign corresponds to \Wp\ (\Wm) boson.

\subsection{Fit procedure applied to Monte Carlo samples}
The fitting procedure with templates was first applied to the simulated samples, at three different levels:
\begin{itemize}
\item all events using generator information for \ctd\ distributions;
\item events remaining after applying acceptance and \trm\ cuts using generator information for \ctd\ distributions;
\item events after the complete event selection (standard plus analysis cuts), using fully simulated information followed by reconstruction 
  for \ctdd\ distributions.
\end{itemize}

The fits of \ctd\ and \ctdd\ distributions were performed using a binned maximum-likelihood fit\ \cite{Barlow1993dm,ROOT}. Since the parameters 
of the fit, \fl, \fL\ and \fR, must sum to 1, only two independent parameters, chosen to be \fl\ and \fLmR, are reported. The parameters were not 
individually constrained to be between 0 and 1.\par
For the second and third steps, numerical results for \fl\ and \fLmR\ fits are summarised in \Tab{MCfitresults} for  35 $<$ \ptw\ $<$ 50 
\GeV\ and \ptw\ $>$ 50 \GeV. In \Tab{MCfitresults} and in the following, the coefficients \fl\ and \fLmR\ represent helicity fractions, averaged 
over \yw, within a given \ptw\ bin.\par
 
\begin{table*}
  \centering
  \caption{Results (as percentages) of fitting \ctd\ and \ctdd\ distributions from \mcnlo\ simulated samples using helicity templates. The fits are
    performed at generator-level, after applying acceptance and \trm\ cuts, and on fully simulated events, after applying standard plus analysis 
    selections using \ctdd. \label{tab:MCfitresults}}
  \begin{tabular}{|c|c|c|c||c|c|}
    \cline{3-6}
    \multicolumn{2}{c|}{} & $\mu^+$ & $\mu^-$ & $e^+$ & $e^-$ \\
    \cline{3-6}
    \multicolumn{2}{c|}{}&\multicolumn{4}{|c|}{35$<$ \ptw\ $<$ 50 \GeV \bigstrut} \\
    \hline
    \multirow{2}{*}{\ctd\ generator-level} & \fl\ $(\%)$& 14.6 $\pm$ 0.8 & 20.9 $\pm$ 0.8 & 15.3 $\pm$ 0.8 & 20.4 $\pm$ 0.9 \\ 
    & \fLmR\ $(\%)$& 27.9 $\pm$ 0.7 & 26.5 $\pm$ 0.8 & 28.2 $\pm$ 0.7 & 26.4 $\pm$ 0.8 \\
    \hline
    \multirow{2}{*}{\ctdd\ fully simulated} & \fl\ $(\%)$& 30.1 $\pm$ 2.4 & 19.5 $\pm$ 2.2 & 26.9 $\pm$ 2.2 & 21.6 $\pm$ 2.3 \\ 
    & \fLmR\ $(\%)$& 31.8 $\pm$ 1.4 & 26.5 $\pm$ 1.2 & 27.3 $\pm$ 1.4 & 22.5 $\pm$ 1.4 \\
    \hline
    \multicolumn{6}{c}{} \\
    \cline{3-6}
    \multicolumn{2}{c|}{}&\multicolumn{4}{|c|}{\ptw\ $\ge$ 50 \GeV \bigstrut} \\
    \hline
    \multirow{2}{*}{\ctd\ generator-level} & \fl\ $(\%)$& 18.3 $\pm$ 1.0 & 22.7 $\pm$ 1.0 & 19.0 $\pm$ 0.9 & 22.1 $\pm$ 1.0 \\ 
    & \fLmR\ $(\%)$& 26.9 $\pm$ 0.8 & 25.8 $\pm$ 0.9 & 27.6 $\pm$ 0.8 & 25.9 $\pm$ 0.9 \\
    \hline
    \multirow{2}{*}{\ctdd\ fully simulated} & \fl\ $(\%)$& 25.1 $\pm$ 1.9 & 20.7 $\pm$ 2.2 & 24.9 $\pm$ 1.8 & 22.5 $\pm$ 2.0 \\ 
    & \fLmR\ $(\%)$& 29.7 $\pm$ 1.1 & 26.2 $\pm$ 1.2 & 25.6 $\pm$ 1.2 & 22.6 $\pm$ 1.3 \\
    \hline
  \end{tabular}
\end{table*}
Template fit results using the \ctd\ distributions at the generator-level, without any cut, reproduce the average value of the numbers 
quoted in the relevant \ptw\ bin of \Fig{ComputedFractions}. With respect to these fit results, the numbers shown in the first lines of 
\Tab{MCfitresults} for the two \ptw\ bins reflect the effect of the acceptance and \trm\ cuts, which is small on \fl\ but is sizeable on 
\fLmR, typically reducing it by 25\% (relative). Indeed, the detector has a small acceptance for the events produced at high \absyw, for 
which \fLmR\ is largest.\par
Comparisons of the first row of each part of \Tab{MCfitresults} (\ctd\ at generator-level, within acceptance) to the second row (\ctdd\ 
after full simulation) indicates that the values of \fl\ are rather stable for \Wm\, while for \Wp\ there is in several cases a 
significant increase. Similar effects are observed with \pow. Corrections applied at the analysis level (see Section\ \ref{sect:systemp}) 
are intended to remove these effects to obtain the final, corrected results.\par

\section{Fit results}
\label{sect:fitresults}

The raw helicity fractions for each of the four analysed channels were obtained by fitting the experimental \ctdd\ distributions, after 
background subtraction, with a sum of templates (see \Eqn{fracdef}) corresponding to longitudinal, left- and right-handed states.\par
In order to correct for systematic effects associated with the choice of the variable used in the fit (\ctdd), and for resolution effects, 
the raw results have been corrected in a second step by the differences observed in Monte Carlo events between the fits at the generator level 
with the \ctd\ distribution after acceptance plus \trm\ cuts and the fit on on \ctdd\ distributions after full simulation. The two sets of 
templates obtained from \mcnlo\ or from \pow\ were used, and their bias corrected for accordingly. Differences between the results obtained with 
the two Monte Carlo generators were used to estimate a systematic uncertainty associated with the choice of templates (see Section\ 
\ref{sect:systemp}).\par 
The minimisation \cite{Barlow1993dm} gives the uncertainties and correlations between the parameters. The $\chi^2$ values, in 
\Tab{chi2valuesfit}, obtained using \mcnlo\ and \pow\ templates, are similar. They are significantly lower, in most cases, than in 
\Tab{chi2values}, especially for muons, even taking into account that the number of degrees of freedom is reduced from 19 to 17.
\begin{table*}
  \centering
  \caption{Values of the $\chi^2$ from the fit of data with \mcnlo\  and \pow\ helicity templates (see \Figs[electronMedPT]{electronHPT} for 
    \mcnlo). The number of degrees of freedom in the fits is 17.}
  \begin{tabular}{|c|c|c|c|c||c|c|c|c|}
    \hline
    \multicolumn{1}{|c|}{\multirow{2}{*}{\minitab[c]{$\chi^2$ between\\ data and}}}& \multicolumn{4}{c||}{35 $<$ \ptw\ $<$50 \GeV \bigstrut} & \multicolumn{4}{c|}{\ptw\ $>$50 \GeV} \\
    \cline{2-9}
    & $\mu^+$ & $\mu^-$ & $e^+$ & $e^-$ & $\mu^+$ & $\mu^-$ & $e^+$ & $e^-$ \\
    \hline
    \mcnlo\  templates & 13.5 & 23.1 & 7.6 & 25.3  & 29.3 & 21.1 & 24.8 & 16.9  \\ 
    \hline
    \pow\  templates & 11.1 & 20.7 & 8.2 & 20.8 & 30.1 & 26.6 & 20.9 & 13.1 \\ 
    \hline 
  \end{tabular}
  \label{tab:chi2valuesfit}
\end{table*}

\begin{figure*}
  \begin{changemargin}{-0.05\textwidth}{-0.05\textwidth}
    \centering
    \subfigure[\mupl]{
      \includegraphics[width=0.38\textwidth]{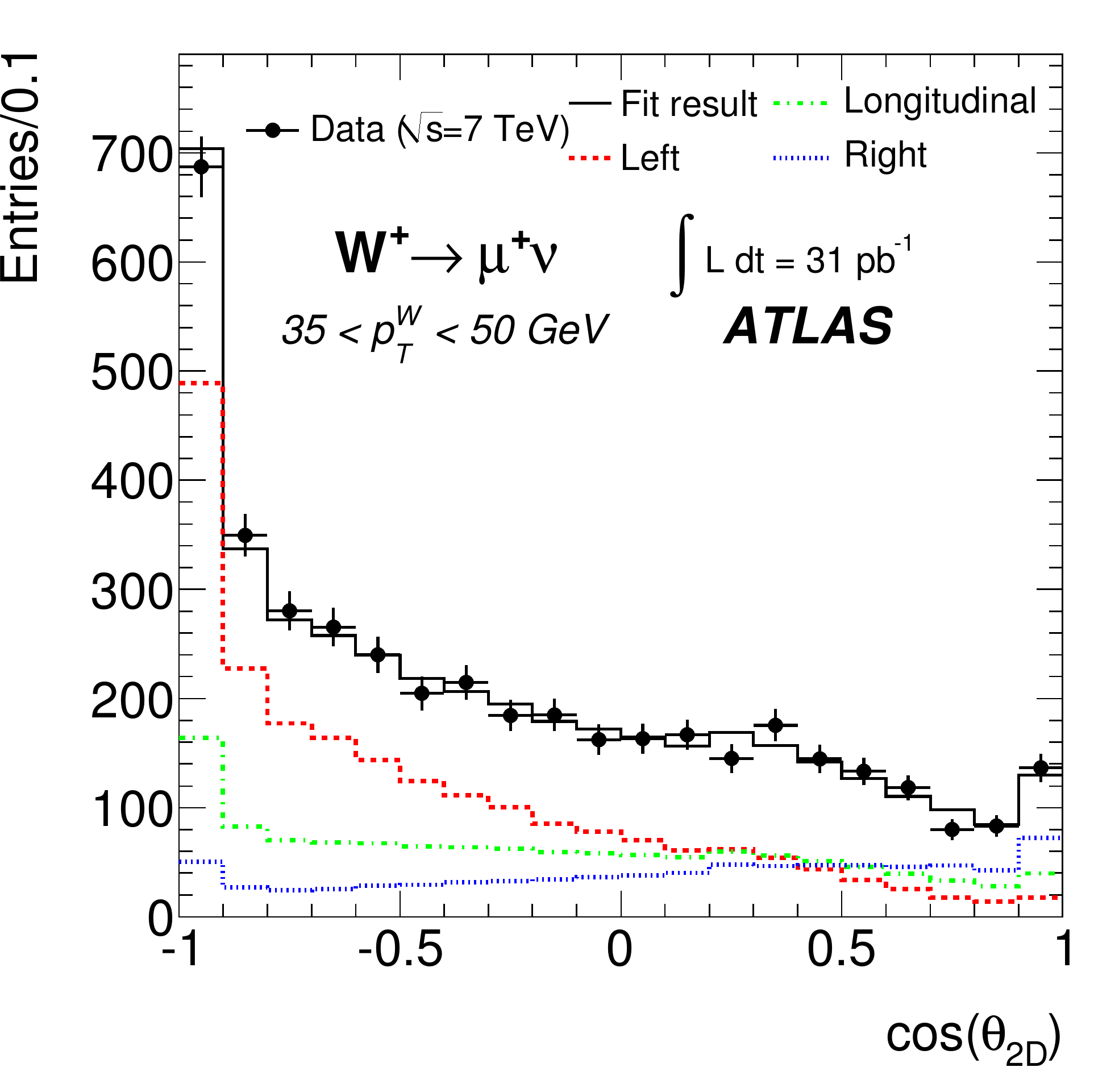}
    }\hspace*{-0.5cm}
    \subfigure[\mumi]{ 
      \includegraphics[width=0.38\textwidth]{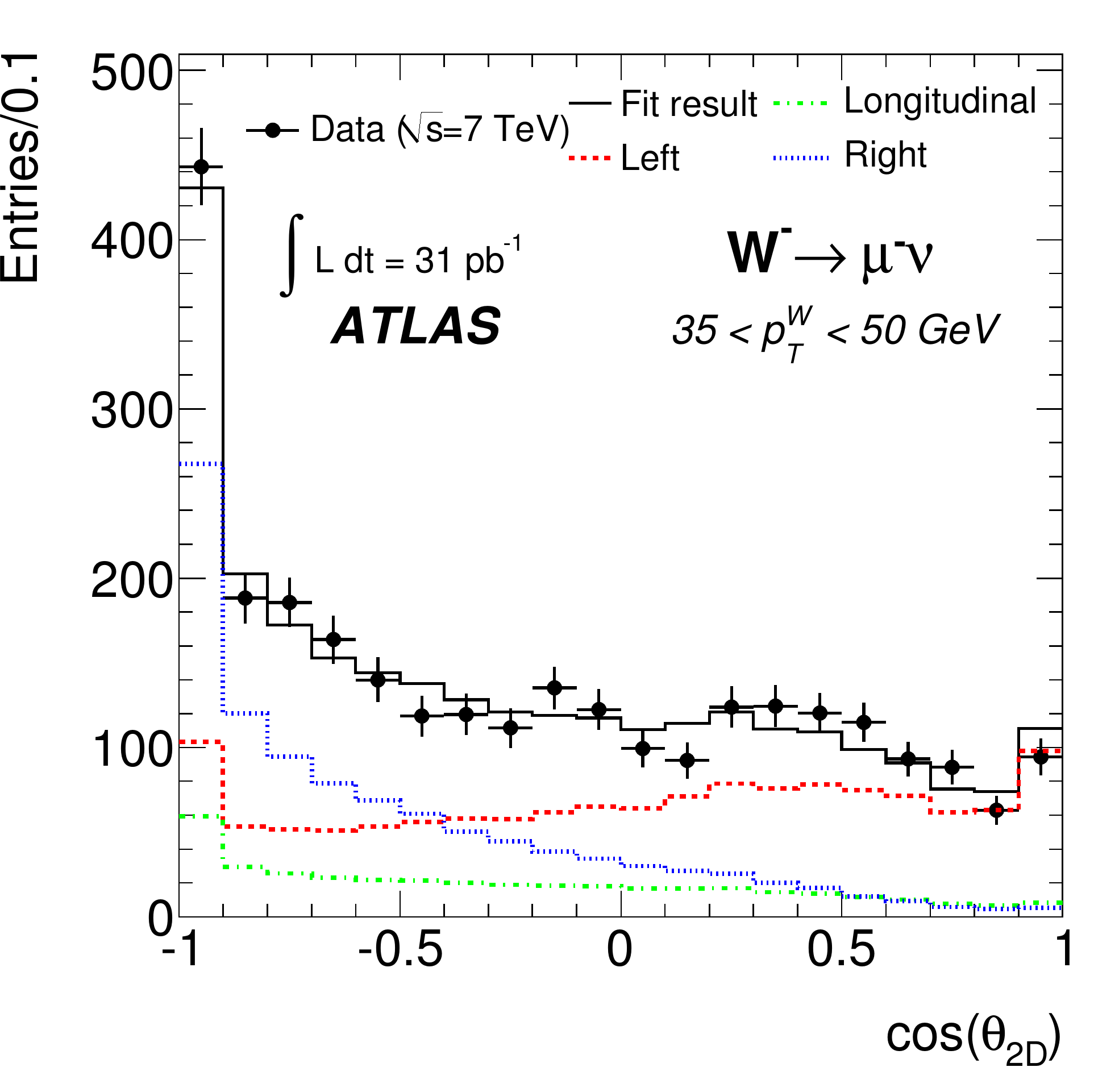}
    }
    \subfigure[\epl]{
      \includegraphics[width=0.38\textwidth]{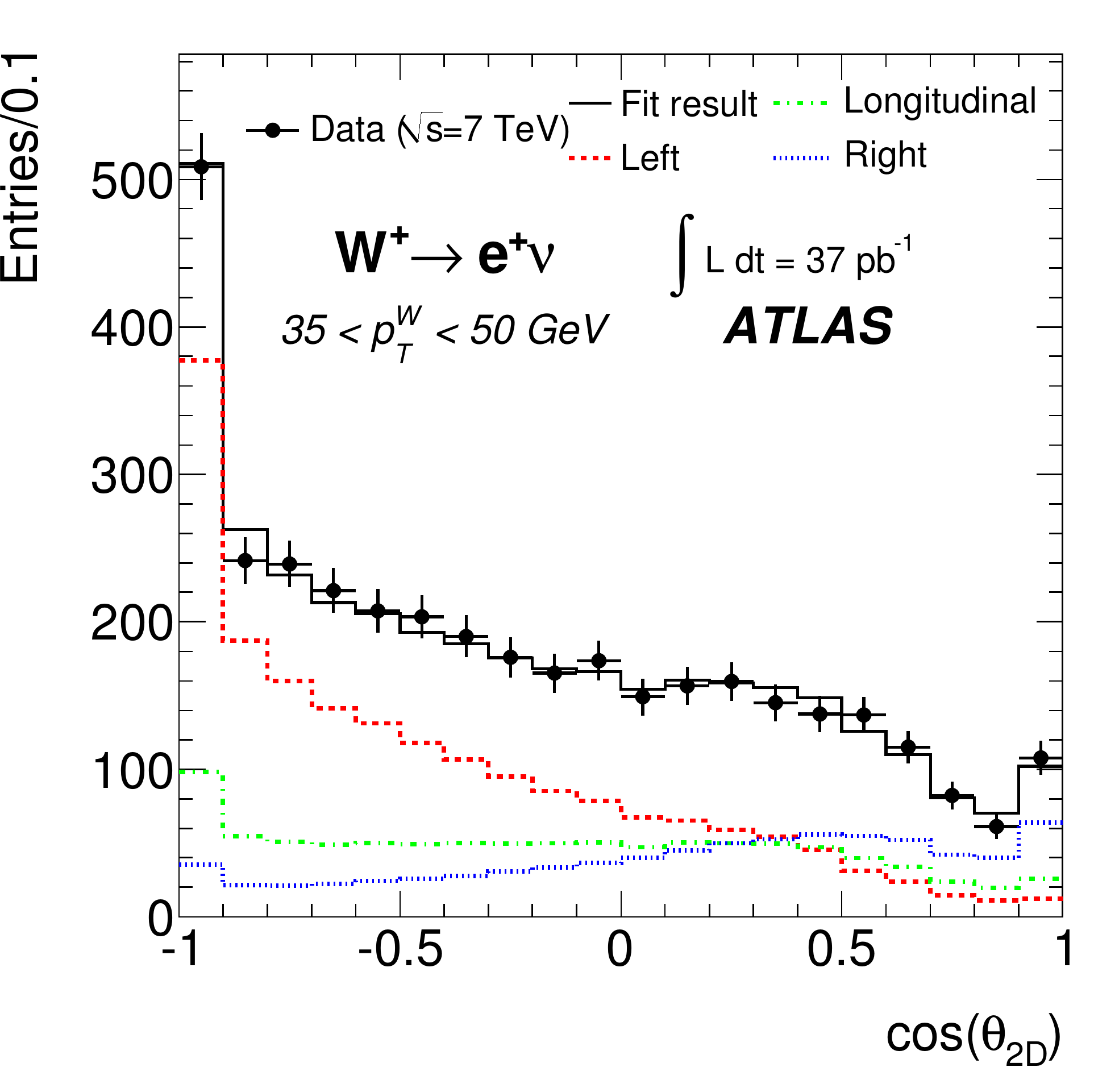}
    }\hspace*{-0.5cm}
    \subfigure[\emi]{
      \includegraphics[width=0.38\textwidth]{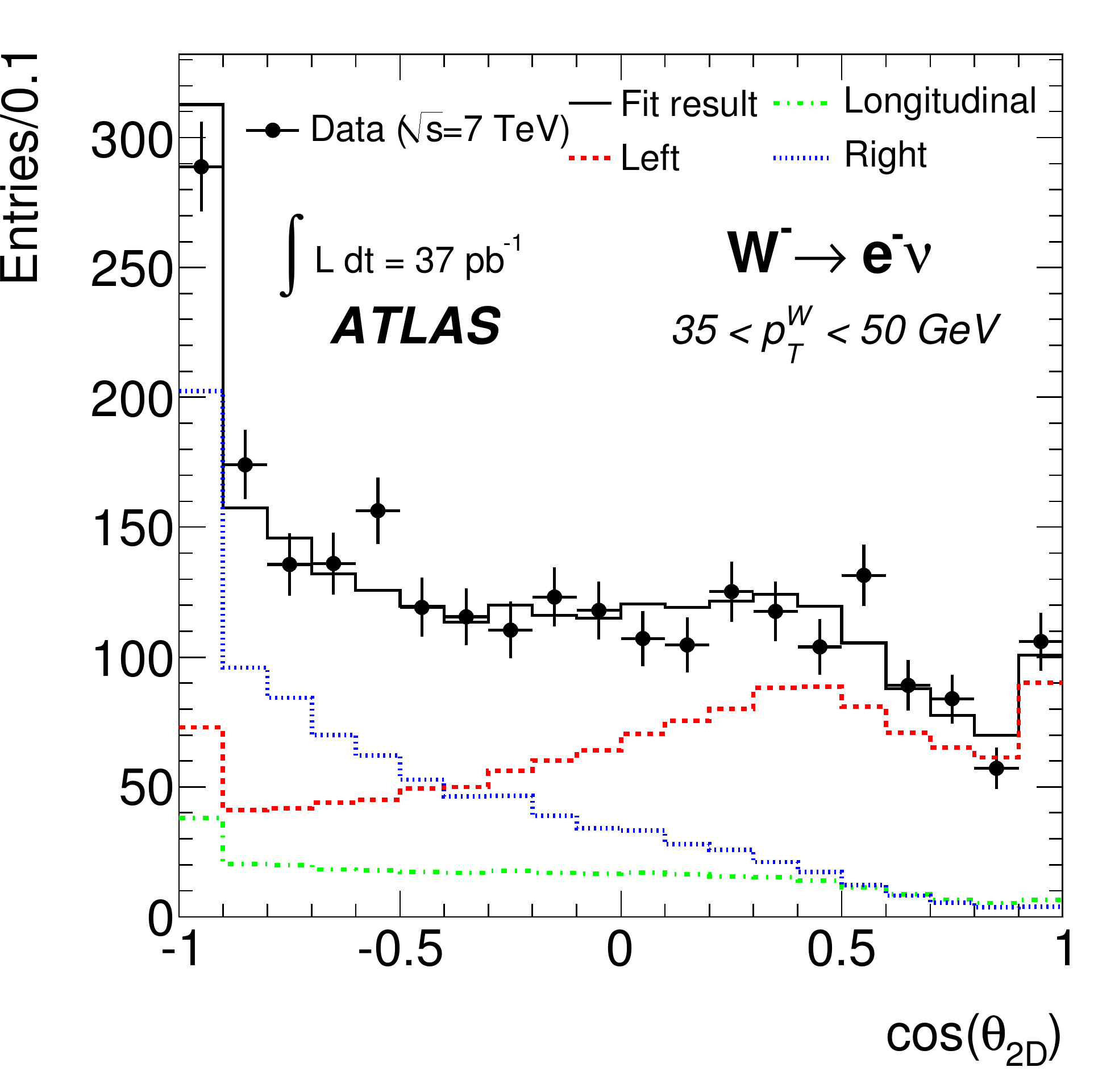}
    }
  \end{changemargin}
  \caption{Results of the fits to \ctdd\ distributions using helicity templates (built from \mcnlo), for \Wmn\ (top) and \Wen\ (bottom) events in 
    data with 35 $<$ \ptw\  $<$ 50 \GeV, after background subtraction. Each template distribution is represented: left-handed contribution (dashed 
    line), longitudinal contribution (dotted-dashed line) and right-handed contribution (dotted line).\label{fig:electronMedPT}}
\end{figure*}

\begin{figure*}
  \begin{changemargin}{-0.05\textwidth}{-0.05\textwidth}
    \centering
    \subfigure[\mupl]{
      \includegraphics[width=0.38\textwidth]{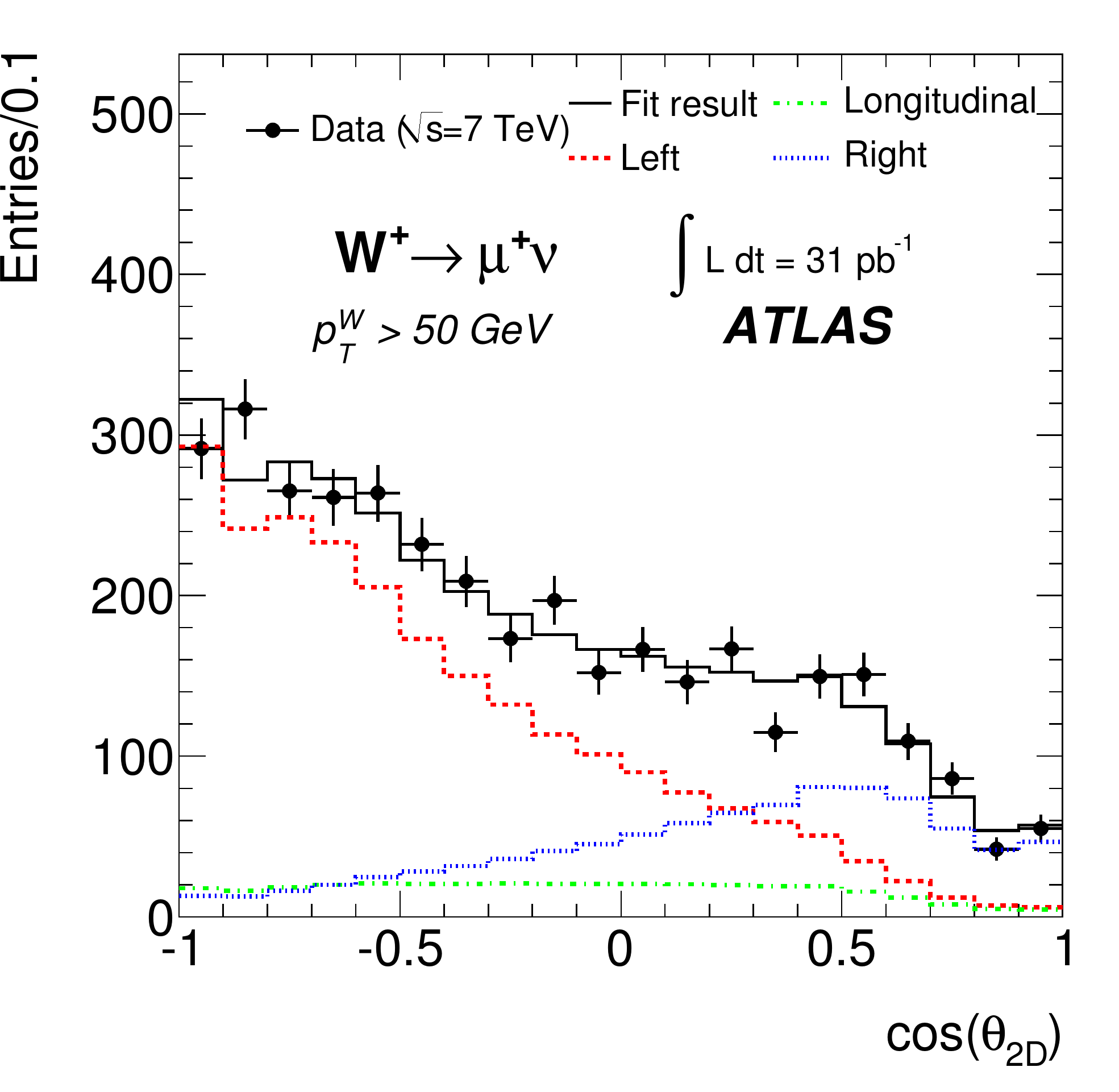}
    }\hspace*{-0.5cm}
    \subfigure[\mumi]{
      \includegraphics[width=0.38\textwidth]{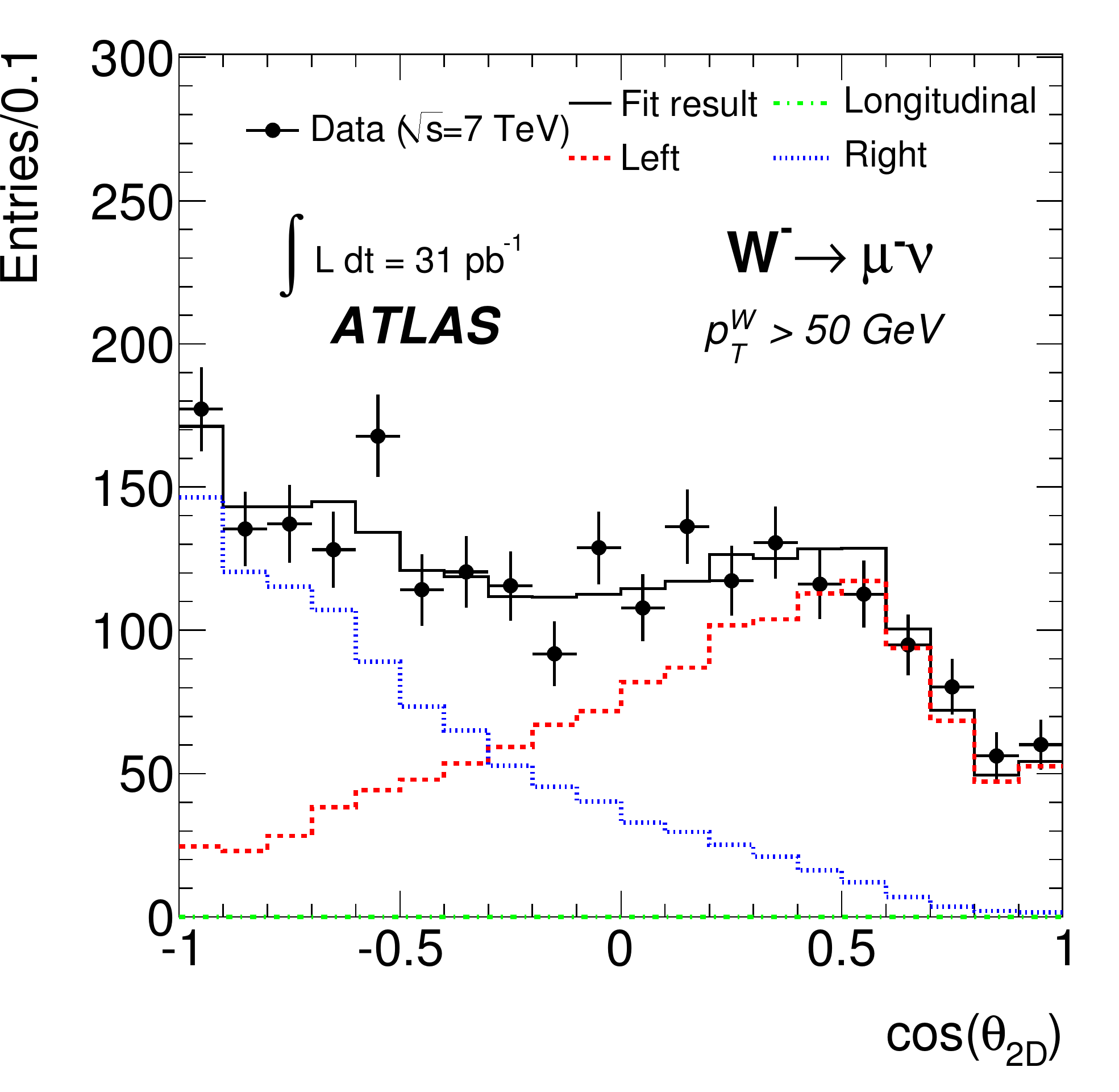}
    }
    \subfigure[\epl]{
      \includegraphics[width=0.38\textwidth]{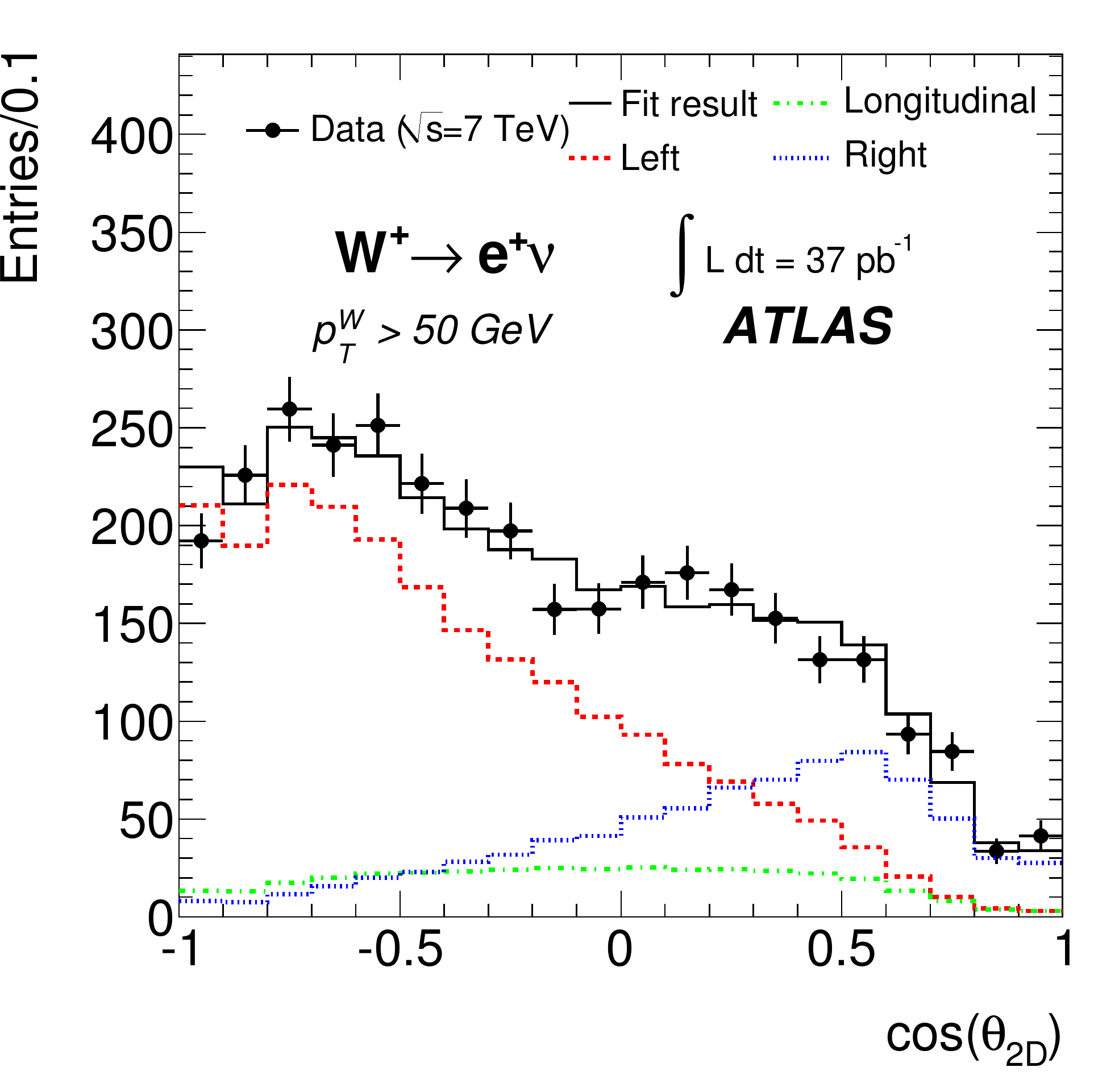}
    }\hspace*{-0.5cm}
    \subfigure[\emi]{
      \includegraphics[width=0.38\textwidth]{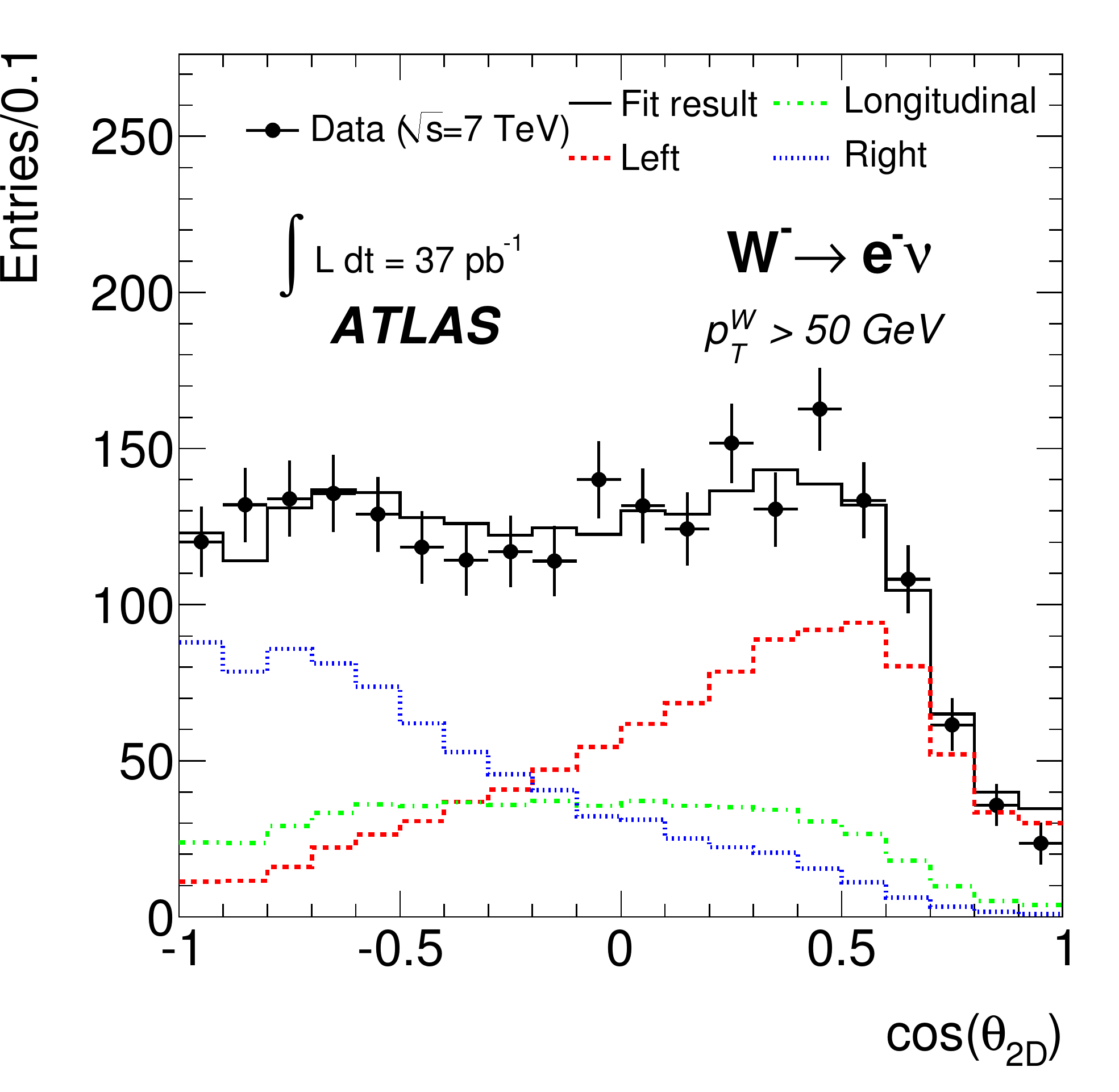}
    }
  \end{changemargin}
  \caption{Results of the fits to \ctdd\ distributions using helicity templates (built from \mcnlo), for \Wmn\ (top) and \Wen\ (bottom) events in 
    data with \ptw\ $>$ 50 \GeV, after background subtraction. Each template distribution is represented: left-handed contribution (dashed 
    line), longitudinal contribution (dotted-dashed line) and right-handed contribution (dotted line).\label{fig:electronHPT} }
\end{figure*}

The values of the fitted parameters, using \mcnlo\  and \pow\  templates, are reported in \Tab{rawdatatot35}. The contributions 
of the individual fitted helicity states, and their sum, are also shown, for the \mcnlo\  case, in \Fig{electronMedPT} for 35 $<$ 
\ptw\ $<$ 50 \GeV, and in \Fig{electronHPT} for \ptw\ $>$ 50 \GeV. These histograms show the contributions of each polarisation 
state (separately and summed together), with a normalisation which, in addition to the value of \fl, \fL\ and \fR, also takes into 
account the relative average acceptance for each of the three polarisation states. The data show a dominance of the left-handed over 
the right-handed fraction in about the same proportion as in the Monte Carlo simulations.\par

\begin{table*}
  \begin{changemargin}{-0.3cm}{-0.3cm}
    \centering
    \footnotesize
    \caption{Summary of raw data results for helicity fractions (as percentages) for 35 $<$ \ptw\ $<$ 50 \GeV\ and \ptw\ $>$ 50 \GeV\ 
      obtained with \mcnlo\ or with \pow\ template fits (see \Figs[electronMedPT]{electronHPT} for \mcnlo). The errors represent the 
      statistical uncertainties only.\label{tab:rawdatatot35}}
    \begin{tabular}{|c|c|c|c||c|c|}
      \cline{3-6}
      \multicolumn{2}{c|}{ }& \mupl & \mumi & \epl & \emi \\
      \cline{3-6}
      \multicolumn{2}{c|}{} & \multicolumn{4}{c|}{35 $<$ \ptw\ $<$ 50 \GeV} \bigstrut \\
      \hline
      \multirow{2}{*}{\minitab[c]{Data \\ with \mcnlo}} & \fl\ $(\%)$ & 26.6 $\pm$ 5.1 & 10.9 $\pm$ 5.6 & 23.2 $\pm$ 5.7 & 9.9 $\pm$ 10.2 \\ 
      & \fLmR\ $(\%)$& 20.6 $\pm$ 3.9 & 27.1 $\pm$ 4.3 & 17.9 $\pm$ 4.2 & 33.0 $\pm$ 4.0 \\  
      \hline  
      \multirow{2}{*}{\minitab[c]{Data \\ with \pow\ }} & \fl\ $(\%)$ & 42.8 $\pm$ 5.1 & 35.1 $\pm$ 5.7 & 36.9 $\pm$ 9.1 & 26.5 $\pm$ 6.1 \\ 
      & \fLmR\ $(\%)$& 25.6 $\pm$ 3.9 & 21.8 $\pm$ 4.3 & 21.3 $\pm$ 5.3 & 25.1 $\pm$ 4.3 \\
      \hline  
      \multicolumn{6}{c}{ } \\
      \cline{3-6}
      \multicolumn{2}{c|}{} & \multicolumn{4}{c|}{\ptw\ $>$ 50 \GeV} \bigstrut \\
      \hline
      \multirow{2}{*}{\minitab[c]{Data \\ with \mcnlo}} & \fl\ $(\%)$& 8.3 $\pm$ 5.0 & -0.0 $\pm$ 7.3 & 9.7 $\pm$ 5.7 & 20.0 $\pm$ 5.6 \\ 
      & \fLmR\ $(\%)$& 27.5 $\pm$ 3.3 & 29.9 $\pm$ 3.4 & 29.3 $\pm$ 3.5 & 19.7 $\pm$ 3.9 \\
      \hline  
      \multirow{2}{*}{\minitab[c]{Data s\\ with \pow\ }} & \fl\ $(\%)$& 15.3 $\pm$ 4.4 & 13.0 $\pm$ 5.0 & 19.6 $\pm$ 5.7 & 26.6 $\pm$ 6.9 \\ 
      & \fLmR\ $(\%)$& 27.7 $\pm$ 3.2 & 19.9 $\pm$ 3.6 & 29.5 $\pm$ 3.6 & 13.3 $\pm$ 4.2 \\
      \hline  
    \end{tabular}
  \end{changemargin}  
\end{table*}

The \fl\ values obtained with the \pow\ templates are in general larger (see \Tab{rawdatatot35}). For the negative charges, the increase 
of \fl\ is correlated with a decrease of \fLmR, while for positive charges the reverse is observed, though with a smaller increase, especially 
in the higher \ptw\ bin.\par

\section{Systematic effects}
\label{sect:systematic}
In addition to the choice of templates, which is treated separately, the measurement suffers from systematic effects due to limited knowledge 
of backgrounds, charge mis-identification, choice of PDF sets, uncertainties on  the lepton energy scale and resolution, and uncertainties on 
the recoil system energy scale and resolution. The uncertainties on helicity fractions have been estimated using \mcnlo\ and are reported in 
\Tab{systtot35}, in absolute terms.\par 
The effect of reweighting simulated  events to restore a \ptw\ distribution closer to that observed \cite{Belloni1361975} was also assessed.

\subsection{Backgrounds}
\label{sect:backnorm}

The electroweak and \ttbar\ backgrounds have been studied previously and found to be well modelled by Monte Carlo simulations\ \cite{Aad2011dm,
Aad2011fu,Aad2011kt,Aad2012qf}. As these backgrounds are subtracted from data for the final fit, an associated systematic uncertainty has been 
estimated by changing the global normalisation of the subtracted distributions by $\pm$ 6.8\% ($\pm$ 3.4\% to take into account the uncertainty 
on the integrated luminosity, $\pm$ 5\% for the uncertainty on background cross-sections relative to signal, and $\pm$ 3\% for the influence of 
PDFs on the acceptance\ \cite{Aad2010yt}).\par
Furthermore, the amount of jet background was varied inside the uncertainty estimated by the dedicated fit (see \Tab{QCDfractions}).\par

\subsection{Charge mis-identification}

Since charge mis-identification is well reproduced by simulations\ \cite{Aad2011mk}, the possible associated effect on the results presented 
here has been measured by comparing helicity fractions extracted from fully simulated events where the charge assignment was taken either from 
ge\-ne\-ra\-tor-level information or after full reconstruction. The effect on \fl\ and \fLmR\ is estimated to be about 0.4\% in the electron case, and 
is negligible for muons.\par

\subsection{Reweighting of \ptw\ distribution}
\label{sect:ptwrew}
\mcnlo\ and, to a lesser extent \pow, underestimate the fraction of \Wboson\ events at high \ptw. In order to investigate the 
possible consequences of such a bias on this measurement, the \mcnlo\  Monte Carlo signal sample, weighted event-by-event so as to  restore a \ptw\ 
spectrum compatible with data, was fitted using unchanged helicity templates (both \pow\ and \mcnlo\ templates were used for this test). The effect 
of the reweighting was found to have a small impact on the fitted values of  \fl\ (less than 2\%). For \fLmR\ sizeable effects were observed (up 
to 5\% in the low \ptw\ bin). However, they are of opposite sign for the positive and negative lepton charges, and almost perfectly cancel when 
analysing charge-averaged values (see \Tab{systtot35}). 

\subsection{PDF sets}

Using the PDF reweighting method, the uncertainty associated with PDFs was estimated by keeping the templates unchanged and using MSTW~2008 and 
HERAPDF~1.0 instead of the CTEQ~6.6 PDFs for the simulation of the signal distributions. The impact on \fl\ and \fLmR\ is in the range of 1\% to 2\%.

\begin{table*}
  \begin{changemargin}{-0.3cm}{-0.3cm}
    \centering
    \footnotesize
    \caption{Summary of systematic uncertainties on helicity fractions for 35 $<$ \ptw\ $<$ 50 \GeV\ and \ptw\ $>$ 
      50 \GeV. The effect of lepton and recoil energy scales, and of \ptw\ reweighting, on \fLmR\ is also estimated on the mean 
      between the two charges. The larger errors appear with the $\pm$ ($\mp$) sign if they vary in the same (opposite) direction 
      as the parameter studied, in order to highlight the correlations used in calculating the errors on the means.\label{tab:systtot35}}
    \begin{tabular}{|c|c|c|c||c|c|c|c|c||c|c|}
      \cline{3-6} \cline{8-11}
      \multicolumn{2}{c|}{} & \multicolumn{4}{c|}{35 $<$ \ptw\ $<$ 50 \GeV} & \multicolumn{1}{c|}{} & \multicolumn{4}{c|}{\ptw\ $>$ 50 \GeV} \bigstrut \\
      \cline{3-6} \cline{8-11}
      \multicolumn{2}{c|}{ }& \mupl & \mumi & \epl & \emi & \multicolumn{1}{c|}{} & \mupl & \mumi & \epl & \emi \\
      \cline{1-6} \cline{8-11}
      \multirow{2}{*}{EW background} & $\delta \fl$ $(\%)$ & 0.5 & 0.6 & 0.3 & 0.4 & \multicolumn{1}{c|}{} & 0.6 & 0.6 & 0.3 & 0.5 \\ 
      & $\delta(\fLmR)$ $(\%)$ & 0.2 & 0.3 & 0.2 & 0.2 & \multicolumn{1}{c|}{} &  0.2 & 0.3 & 0.2 & 0.2 \\
      \cline{1-6} \cline{8-11}
      \multirow{2}{*}{jet background} & $\delta \fl$ $(\%)$ & 1.5 & 1.5 & 1.5 & 1.5  & \multicolumn{1}{c|}{} &  2.3 & 1.3 & 2 & 2 \\ 
      & $\delta(\fLmR)$ $(\%)$ & 0.3 & 0.7 & 1.5 & 1.5  & \multicolumn{1}{c|}{} &  1.2 & 1.3 & 1.5 & 1.5 \\ 
       \cline{1-6} \cline{8-11}
      \multirow{3}{*}{\ptl\ scale} & $\delta \fl$ $(\%)$ & $\mp$ 4.5 & $\mp$ 5.0 & $\mp$ 4.5 & $\mp$ 4.5  & \multicolumn{1}{c|}{} & $\mp$ 3.5  & $\mp$ 3.5 & $\mp$ 3.5 & $\mp$ 4.5 \\
      & $\delta(\fLmR)$ $(\%)$ & $\mp$ 2.5  & $\pm$ 2.0 & $\mp$ 2.5 & $\pm$ 2.0  & \multicolumn{1}{c|}{} & $\mp$ 1.5  & $\pm$ 1.5 & $\mp$ 2.0 & $\pm$ 1.5 \\
      \cline{3-6} \cline{8-11}
      & $\delta(\fLmR)_{\rm mean}$ $(\%)$ & \multicolumn{2}{c||}{1.1} & \multicolumn{2}{c|}{0.4}  & \multicolumn{1}{c|}{} & \multicolumn{2}{c||}{0.1} & \multicolumn{2}{c|}{0.4} \\
       \cline{1-6} \cline{8-11}
      \multirow{3}{*}{Recoil scale} & $\delta \fl$ $(\%)$ & $\pm$ 12.5  & $\pm$ 16.8 & $\pm$ 12.5  & $\pm$ 13.3  & \multicolumn{1}{c|}{} & $\pm$ 8.1  & $\pm$ 10.2 & $\pm$ 9.4  & $\pm$ 11.1 \\
      & $\delta(\fLmR)$ $(\%)$ & $\pm$ 9.9 & $\mp$ 10.4 & $\pm$ 10.9 & $\mp$ 9.5  & \multicolumn{1}{c|}{} & $\pm$ 7.7 & $\mp$ 7.7 & $\pm$ 8.2 & $\mp$ 8.2 \\
      \cline{3-6} \cline{8-11}
      & $\delta(\fLmR)_{\rm mean}$ $(\%)$ & \multicolumn{2}{|c||}{3.0} & \multicolumn{2}{c|}{2.9}  & \multicolumn{1}{c|}{} & \multicolumn{2}{c||}{1.2} & \multicolumn{2}{c|}{0.7} \\
       \cline{1-6} \cline{8-11}
      \multirow{2}{*}{PDF set} & $\delta \fl$ $(\%)$ & 2.0  & 2.0 & 0.4 & 0.8  & \multicolumn{1}{c|}{} & 2.0  & 2.0 & 0.2 & 0.8 \\
      & $\delta(\fLmR)$ $(\%)$ & 1.5  & 1.5  & 0.5 & 1.5  & \multicolumn{1}{c|}{} & 1.5  & 1.5  & 0.4 & 1.1 \\
       \cline{1-6} \cline{8-11}
      \multirow{2}{*}{Charge mis-ID} & $\delta \fl$ $(\%)$ & \multirow{2}{*}{ $-$ }  & \multirow{2}{*}{ $-$ }  & 0.2 & 0.4  & \multicolumn{1}{c|}{} &  \multirow{2}{*}{ $-$ }  & \multirow{2}{*}{ $-$ }  & 0.2 & 0.2 \\
      & $\delta(\fLmR)$ $(\%)$ &   &   & 0.3 & 0.4  & \multicolumn{1}{c|}{} &    &   & 0.2 & 0.3 \\
       \cline{1-6} \cline{8-11}
      \multirow{2}{*}{\ptl\ resolution} & $\delta \fl$ $(\%)$ & 0.1  & 0.1 & 0.5 & 0.5  & \multicolumn{1}{c|}{} & 0.1  & 0.2 & 0.2 & 1.2 \\
      & $\delta(\fLmR)$ $(\%)$ & 0.1  & 0.1  & 0.3  & 0.3  & \multicolumn{1}{c|}{} & 0.1  & 0.2  & 0.2  & 0.2 \\
      \cline{1-6} \cline{8-11}
      \multirow{3}{*}{\ptw\ reweighting} & $\delta \fl$ $(\%)$ & 2.5  & 1.1 & 0.6 & 0.9  & \multicolumn{1}{c|}{} & 1.9  & 1.6 & 0.5 & 1.2 \\
      & $\delta(\fLmR)$ $(\%)$ & $\mp$ 4.9  & $\pm$ 5.2  & $\mp$ 4.2  & $\pm$ 4.0  & \multicolumn{1}{c|}{} & $\mp$ 2.7  & $\pm$ 2.9  & $\mp$ 2.6  & $\pm$ 2.3 \\
      \cline{3-6} \cline{8-11}
      & $\delta(\fLmR)_{\rm mean}$ $(\%)$ & \multicolumn{2}{|c||}{0.2} & \multicolumn{2}{c|}{0.1}  & \multicolumn{1}{c|}{} & \multicolumn{2}{c||}{0.1} & \multicolumn{2}{c|}{0.2} \\
      \cline{1-6} \cline{8-11}
    \end{tabular}
  \end{changemargin}  
\end{table*}

\subsection{Energy scales}

\label{sect:escale}
While a coherent change of the lepton and recoil energy scales would leave the angles in the transverse plane unchanged, both in the laboratory 
and in the transverse \Wboson\ rest frame, an effect on \ctdd\ arises when only one of the two measured objects (lepton, recoil) changes, or if 
they change by different amounts.\par
Using simulated events, it has been observed that an increase of the lepton transverse momentum alone gives a positive slope to the \ctdd\ 
distribution, which in turn induces an increase of the left-handed fraction in the negative lepton sample, and a decrease of the left-handed 
fraction in the positive lepton sample. As expected, the reverse happens for an increase of the recoil transverse energy.\par  

The value of \fLmR\ when averaged over the two charges is largely independent of the lepton and recoil energy scales, as can be seen in 
\Tab{systtot35}.\par 
The same compensation mechanism is however not present for \fl, for which an increase in the recoil energy scale induces an increase of \fl\ for 
both charges.\par
The lepton energy scale is precisely determined from \Zll\ decays: using the precisely-known value of the \Zboson\ boson mass, scale factors 
have been extracted by \etal\ regions, which in the muon case depend also on the muon charge\ \cite{Aad2011mk,muonscale}. The reconstructed 
\Zboson\ boson mass spectrum has also been used to derive smearing corrections to be applied to Monte Carlo electrons and muons in order to 
reproduce the observed \Zboson\ mass peak resolution. The resulting uncertainties are about 3\% to 5\% on \fl\ and around 2\% on \fLmR.\par
For the rather large \pt\ of the  \Wbosons\ studied here, the recoil system in general contains one or several jets with \pt\ $>$ 20 \GeV, and may 
also include additional ``soft jets'' (7 $<$ \pt\ $<$ 20 \GeV), and clusters of calorimeter cells not included in the above objects. The uncertainty 
on the energy scale of these objects (typically 3\% for jets, 10.5\% for soft jets and 13.5\% for isolated clusters) was propagated as described in\ 
\cite{Aad2011re}. This is the largest systematic uncertainty on the helicity fractions measured in this study. In the worst case (muons in the low 
\ptw\ bin), the resulting uncertainty on \fl\ is 16\%. This uncertainty is largely correlated between the muon and electron channels.\par
Given the anti-correlation observed between the impacts on positive and negative leptons, the uncertainties from energy scale variations enter with 
$\pm$\ or $\mp$\ in \Tab{systtot35}, depending on whether the effect goes in the same direction as an energy increase or in the opposite direction. As 
already pointed out, in the case of \fLmR\ the effects largely cancel when considering the average between negative and positive charges.\par

\subsection{Choice of the Monte Carlo generator}
\label{sect:systemp}

The results of the template fits to real and fully simulated data are affected by the imperfect correlation between \ctdd\  and \ctd\  and by 
resolution effects.\par
In order to compare results directly to theoretical models, the raw results from Section\ \ref{sect:fitresults} are corrected by adding the 
difference, found using simulations, between the ``true'' values which would be given by fits to \ctd\  distributions obtained at the generator 
level within acceptance and \trm\ cuts as used here, and the results obtained using fully-simulated \ctdd\ distributions. In order to be able to 
average results from muons and electrons, the electron results are corrected to the same \etal\ acceptance as for muons (i.e. without  
the barrel-endcap calorimeters overlap region around 1.5, and with a maximum $|\etal|$ value of 2.4).\par
The corrections for results obtained using \mcnlo\  templates were determined from the difference between:
\begin{itemize}
\item results of a fit of \mcnlo\ (3D) templates to \ctd\ distributions of the \pow\ Monte Carlo samples at the generator-level with acceptance 
  and \trm\ cuts.
\item results of a fit of  \mcnlo\ (2D) templates to \ctdd\ distributions of the same \pow\ Monte Carlo samples, after full simulation and with standard 
  plus analysis cuts.
\end{itemize}
The corrections for results obtained using \pow\ templates were derived in the same way as above, interchanging the roles of  \mcnlo\  and \pow.

\begin{table}
  \centering
  \caption{Percentage values of \fLmR\ and \fl\ averaged over charges, separately for electrons and muons, obtained by averaging results with 
    templates from \mcnlo\  (see \Figs[electronMedPT]{electronHPT}) and from \pow. The first uncertainty is statistical, the second covers the 
    systematic uncertainties from instrumental and analysis effects, and the last one the differences between templates constructed with the two 
    generators.\label{tab:correctedfLmR}}
  \begin{changemargin}{-0.3cm}{0cm}
    \begin{tabular}{|c|c||c|}
      \cline{2-3}
      \multicolumn{1}{c|}{}& 35$<$\ptw$<$50 \GeV & \ptw$>$50 \GeV \bigstrut\\ 
      \cline{2-3}
      \multicolumn{1}{c|}{}& \multicolumn{2}{c|}{ \fLmR\ $(\%)$} \bigstrut\\ 
      \hline
      muon average & 21.7$\pm$3.0$\pm$3.6$\pm$2.0 & 25.0$\pm$2.5$\pm$2.3$\pm$2.5 \\
      \hline
      electron average & 26.0$\pm$2.8$\pm$3.4$\pm$2.0 & 25.5$\pm$2.6$\pm$2.0$\pm$2.0 \\
      \hline
      \multicolumn{3}{c}{} \\
      \cline{2-3} 
      \multicolumn{1}{c|}{}& \multicolumn{2}{c|}{ \fl\ $(\%)$ } \bigstrut\\ 
      \hline
      muon average & 23.6$\pm$3.8$\pm$12.0$\pm$7.2 & 7.6$\pm$4.8$\pm$9.0$\pm$5.2 \\ 
      \hline
      electron average & 20.1$\pm$6.9$\pm$12.0$\pm$5.0 & 17.7$\pm$4.3$\pm$9.0$\pm$6.0 \\
      \hline
    \end{tabular}
  \end{changemargin}

\end{table}

In a further step, after averaging over the charges for each lepton flavour:
\begin{itemize}
\item the corrected data result, for \fLmR\ and \fl, was obtained by averaging the numbers obtained with \mcnlo\ and with \pow\ templates.
\item the systematic uncertainty associated with the choice of templates  was taken as half the difference between the two numbers, with
  a minimum value of 2\%.
\end{itemize}

The corrected results and the associated systematic uncertainties are shown in \Tab{correctedfLmR} for \fLmR\ and \fl.\par
The systematic uncertainty associated with the differences between the two sets of templates is large for \fl, for 
which other systematic effects are also large.\par
Another correction procedure was tried, using the same Monte Carlo generator for producing the templates and calculating the corrections. 
The resulting central values of the helicity fractions are very close to those shown in \Tab{correctedfLmR} (within less than 2\%), but the systematic
uncertainties of the corrections are slightly larger (by about 10\% in relative terms).\par
Finally, a full simulation based on \sher~1.2.2\ \cite{sherpa}, made only for the electron channel, was also used to obtain, similarly as 
above, first raw results, and then correction terms found by applying \sher\ templates to simulated data produced with both \mcnlo\ and \pow. The 
corrected measurement obtained in this way are shown in \Tab{sherresults}, together with the ``electron average'' results from \Tab{correctedfLmR}. In 
the case of \sher, only the uncertainty associated with the choice of template is reported. A very good agreement is observed.

\begin{table}[h!]
  \footnotesize
  \caption{Corrected values of \fLmR\ and \fl\ (as percentages) obtained using \sher\ templates, compared to the standard result (\Tab{correctedfLmR}), 
    for the electron channels averaged over charges. In the \sher\ case the only uncertainty quoted is associated with the two ways of calculating the 
    correction term: applying \sher\ templates either to \mcnlo\ or to \pow\ simulated data.\label{tab:sherresults}} 
  \begin{changemargin}{-0.6cm}{-0.6cm}
    \centering
    \begin{tabular}{|c|c||c|}
      \cline{2-3}
      \multicolumn{1}{c|}{}& \multicolumn{1}{c|}{35 $<$ \ptw\ $<$ 50 \GeV} & \ptw\ $>$ 50\GeV\bigstrut \\
      \cline{2-3}
      \multicolumn{1}{c|}{} & \multicolumn{2}{c|}{\fLmR\ $(\%)$}  \\
      \hline
      Data (\sher) &  25.5$\pm$2.2 & 26.6$\pm$2 \\
      \hline
      Data (standard) & 26.0$\pm$2.8$\pm$3.4$\pm$2.0 & 25.5$\pm$2.6$\pm$2.0$\pm$2.0 \\
      \hline
      \multicolumn{3}{c}{}\\
      \cline{2-3}
      \multicolumn{1}{c|}{} & \multicolumn{2}{c|}{\fl\ $(\%)$}  \\
      \hline
      Data (\sher) &  21.0 $\pm$ 9.1  & 15.6 $\pm$ 6.1 \\
      \hline
      Data (standard) & 20.1$\pm$6.9$\pm$12.0$\pm$5.0 & 17.7$\pm$4.3$\pm$9.0$\pm$6.0 \\
      \hline
    \end{tabular}

  \end{changemargin}
\end{table}
\vspace*{-0.3cm}

\section{Results}
\label{sect:final}

The corrected final measurements of \fLmR, already shown in \Tab{correctedfLmR}, are compared in \Tab{fLmRaverage35} to the values obtained from the 
\mcnlo\  and \pow\  samples, at the generator-level with the acceptance and \trm\ cuts, using a template fit to the \ctd\ distributions.\par
In the low \pt\ bin the data lie in between the \mcnlo\  and \pow\  predictions, slightly closer to the former. For \ptw\ $>$ 50 
\GeV, the data are close to the \mcnlo\  values, while \pow\ predicts a somewhat smaller difference between left- and right-handed states than 
observed in the data.\par
The same good agreement between data and \mcnlo\ remains after averaging results over lepton flavours (\Tab{fLRaverage}). While the complete NNLO 
cross-section calculation of Ref. \cite{Bern2011ie} has not been implemented in a Monte Carlo generator, it can be seen in \Fig{ComputedFractions} 
and its equivalent (not shown) for BlackHat, that at the particle level, without any cuts, the \fLmR\ values from\ \cite{Bern2011ie} are on average 
about 5\% lower (in absolute terms) than the \mcnlo\ predictions. They are thus quite close to \pow\ and somewhat lower than the data.\par
The measurements shown in \Tab{fLRaverage}, where all systematic  uncertainties have been combined, are the main result of this study concerning \fLmR, 
and the directly related coefficient $A_4$ (\Eqn{a4rel}).\\

\begin{table*}
  \centering

  \caption{Corrected values, of \fLmR\ (as percentages), averaged over charge, separately for electrons and muons, for the data, \mcnlo\  and \pow, 
    and for 35 $<$ \ptw\ $<$ 50 \GeV\ and \ptw\ $>$ 50 \GeV. For data the first uncertainty is statistical, the second covers 
    the systematic uncertainties from instrumental and analysis effects, and the last one the differences between templates constructed with the two 
    generators.  For \mcnlo\ and \pow\ the uncertainties are only statistical.\label{tab:fLmRaverage35}}

  \begin{tabular}{|c|c||c|c|c||c|}
    \cline{2-3} \cline{5-6}
    \multicolumn{1}{c|}{} & \multicolumn{2}{c|}{35 $<$ \ptw\ $<$ 50 \GeV} \bigstrut & \multicolumn{1}{c|}{} & \multicolumn{2}{c|}{\ptw\ $>$ 50 \GeV} \\
    \cline{2-3} \cline{5-6}
    \multicolumn{1}{c|}{} & muon average & electron average & \multicolumn{1}{c|}{} &  muon average & electron average \\
    \cline{1-3} \cline{5-6}
    \multirow{1}{*}{Data } & 21.7 $\pm$ 3.0 $\pm$ 3.6 $\pm$ 2.0 &26.0 $\pm$ 2.8 $\pm$ 3.4 $\pm$ 2.0& \multicolumn{1}{c|}{} &25.0  $\pm$ 2.5 $\pm$ 2.3 $\pm$ 2.5& 25.5 $\pm$ 2.6 $\pm$ 2.0 $\pm$ 2.0\\
    \cline{1-3} \cline{5-6}
    \multirow{1}{*}{\mcnlo\ } & 27.2 $\pm$ 0.8& 27.1 $\pm$ 1.0 & \multicolumn{1}{c|}{} & 26.4 $\pm$ 0.8& 26.1 $\pm$ 0.9 \\ 
    \cline{1-3} \cline{5-6}
    \multirow{1}{*}{\pow\ } & 19.9$\pm$ 0.8&  19.9 $\pm$ 1.0 & \multicolumn{1}{c|}{} & 21.2 $\pm$ 0.8& 21.2 $\pm$ 0.9 \\ 
    \cline{1-3} \cline{5-6}
    \multicolumn{6}{c}{}\\
  \end{tabular}

  \caption{Corrected values of \fLmR\ and \fl\ (as percentages), averaged over charges and lepton flavours, for the data, \mcnlo\ and \pow, and for 
    35 $<$ \ptw\ $<$ 50 \GeV\ and \ptw\ $>$ 50 \GeV (\Fig{Final}). For data the first uncertainty is statistical, the second covers all systematic 
    uncertainties. For \mcnlo\ and \pow\ the uncertainties are only statistical.\label{tab:fLRaverage}}

  \begin{tabular}{|c|c||c|c|c||c|}
    \cline{2-3} \cline{5-6}
    \multicolumn{1}{c|}{} & \multicolumn{2}{c|}{\fLmR\ $(\%)$} &  \multicolumn{1}{c|}{}  & \multicolumn{2}{c|}{\fl\ $(\%)$} \\
    \cline{2-3} \cline{5-6}
    \multicolumn{1}{c|}{}& 35 $<$ \ptw\ $<$ 50 \GeV & \ptw\ $>$ 50 \GeV & \multicolumn{1}{c|}{} & 35 $<$ \ptw\ $<$ 50 \GeV & \ptw\ $>$ 50 \GeV \bigstrut \\
    \cline{1-3} \cline{5-6}
    \multirow{1}{*}{Data} & 23.8  $\pm$ 2.0 $\pm$ 3.4 & 25.2 $\pm$ 1.7 $\pm$ 3.0 & \multicolumn{1}{c|}{} &  21.9  $\pm$ 3.3 $\pm$ 13.4 &12.7 $\pm$ 3.0 $\pm$ 10.8 \\
    \cline{1-3} \cline{5-6} 
    \multirow{1}{*}{\mcnlo\ } & 27.1 $\pm$ 0.7& 26.2 $\pm$ 0.5    & \multicolumn{1}{c|}{} & 17.9 $\pm$ 1.2& 21.0 $\pm$ 1.0    \\
    \cline{1-3} \cline{5-6} 
    \multirow{1}{*}{\pow\ } & 19.9 $\pm$ 1.0& 21.2 $\pm$ 0.8 & \multicolumn{1}{c|}{} & 22.9 $\pm$ 1.0& 19.4 $\pm$ 0.8 \\
    \cline{1-3} \cline{5-6} 
    \multicolumn{6}{c}{} 
  \end{tabular}
  \captionsetup{type=figure}
  \centering
  \includegraphics[width=0.48\textwidth]{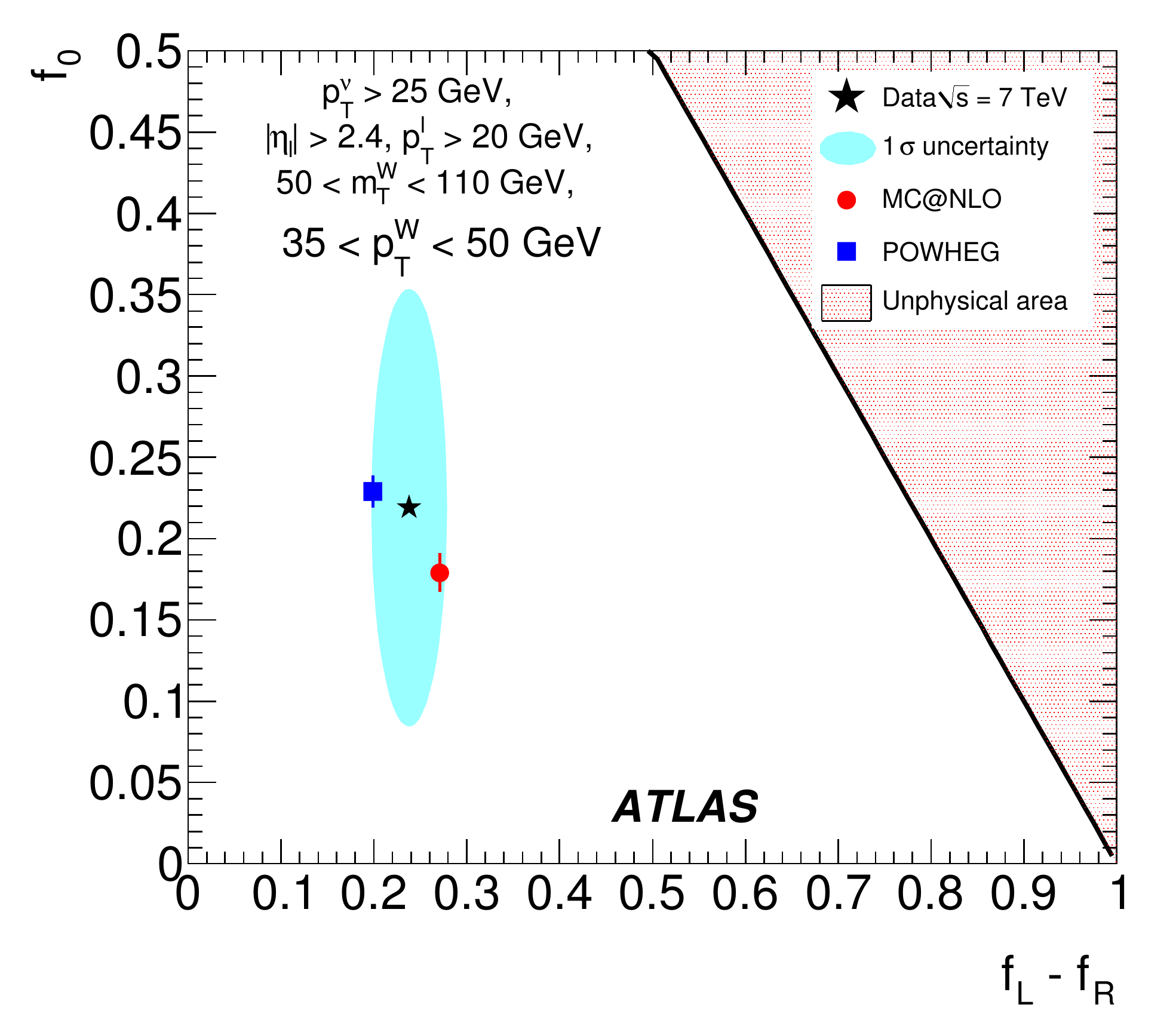}
  \includegraphics[width=0.48\textwidth]{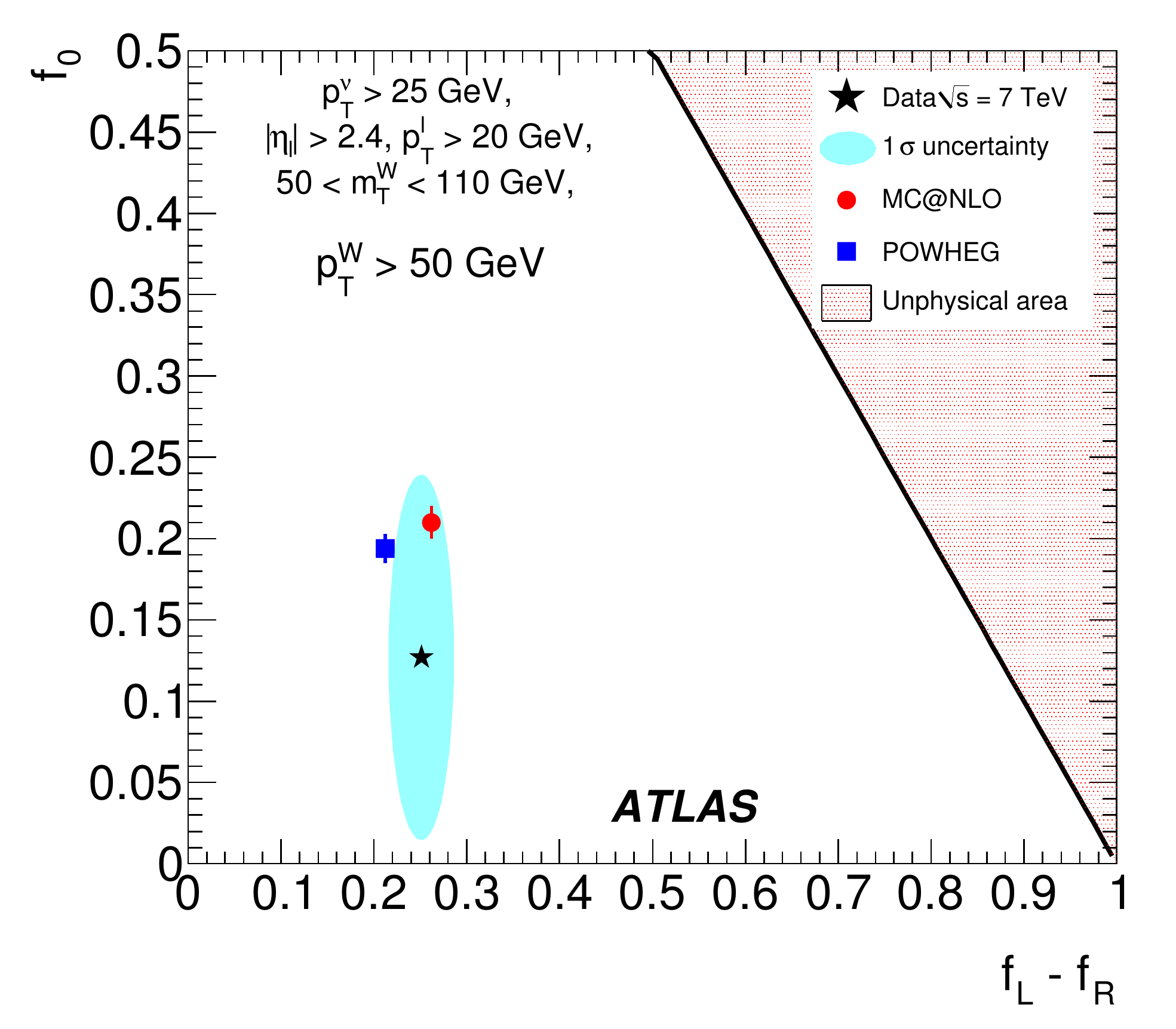}
  \caption{Measured values of \fl\ and \fLmR after corrections (\Tab{fLRaverage}), within acceptance cuts, for 35 $<$ \ptw\ $<$ 50 \GeV\ (left) 
    and \ptw\ $>$ 50 \GeV\ (right), compared with the predictions of \mcnlo\ and \pow. The ellipses around the data points correspond to one 
    standard deviation.\label{fig:Final}}
\end{table*}

For \fl, and the directly related coefficient $A_0$ (\Eqn{fracAi}), the systematic uncertainties associated with the recoil and lepton energy scales 
do not cancel between negative and positive charges. In order to reduce the statistical uncertainties, which are also large, and the uncorrelated 
instrumental and analysis systematic uncertainties, the measurements in each \ptw\ bin were averaged over charges and lepton flavours. The uncertainties 
from the recoil energy scale were taken to be fully correlated among all four measurements. The uncertainty associated with the template model 
(\Tab{correctedfLmR}) was combined quadratically with the other systematic uncertainties.\par

A comparison between the corrected experimental results and the predicted values, within the acceptance and \trm\ cuts (\Tab{fLRaverage}), 
indicates that:
\begin{itemize}
\item
  in the low \ptw\ bin the data are compatible with both  \mcnlo\ and \pow\  predictions, which are mutually consistent. 
\item 
  in the high \ptw\ bin, the data favour \fl\ values smaller than the predictions of \mcnlo\  and \pow, which are close to each other. 
\end{itemize}
Due to the large uncertainties on the measurements, however, no stringent constraints nor clear inconsistencies can be deduced.
The measured values of \fl\ and \fLmR\ are plotted in \Fig{Final} within the triangular region allowed by the constraint \fL+\fl+\fR=1, 
together with the predictions from \mcnlo\ and \pow.\par

\section{Summary and conclusions}
\label{sect:conclusions}
The results presented in this paper show that \mcnlo\  and \pow\  reproduce well the shape of the angular distributions in the transverse 
plane of charged leptons from high-\pt\ \Wboson\ boson decays (\ptw\ $>$ 35 \GeV), a regime where the leading-quark effect in quark-antiquark 
annihilation is subordinate to the dynamics of quark-gluon interactions producing \Wbosons.\par
The variable used for the analysis in terms of helicity fractions (respectively \fl, \fL\ and \fR) is the cosine of the ``transverse helicity'' 
angle \ctdd. Given that the three helicity fractions are  constrained to sum to unity, the independent variables chosen in this study are \fl\ 
and \fLmR. Their values have been derived by fitting \ctdd\ distributions with templates representing longitudinal, left- and right-handed 
\Wbosons. Two sets of templates were used, obtained from \mcnlo\ and \pow.\par
The experimental results have been corrected for the difference between the distribution of the measured quantity, the ``transverse helicity'' 
angle \ctdd, and the distribution of the true helicity angle, \ctd. The correction includes resolution effects, as well as systematic differences 
between the two sets of templates. Corrected results correspond to the following acceptance region: $|\etal|$ $<$ 2.4, \ptn\ $>$ 25 \GeV, \ptl\ $>$ 
20 \GeV\ and 50 $<$ \trm\ $<$ 110 \GeV.\par
The longitudinal fraction is the most difficult to extract and has rather large systematic uncertainties, especially in the low \ptw\ bin, mostly 
associated with the recoil energy scale and with the choice of Monte Carlo generator. In the low \ptw\ bin the data are compatible with both 
\mcnlo\ and \pow\ predictions while in the high \ptw\ bin, they favour lower values than predicted by either of the simulations, which agree well 
with each other.\par
When averaging over charges, \fLmR\ is measured with a small statistical uncertainty and a relatively small systematic uncertainty. The agreement 
between data and \mcnlo, separately for the four measurements (two lepton flavours and two\ \ptw\ bins) is good. Predictions by \pow\  are  somewhat 
smaller than data, especially in the high \ptw\ bin.\par

\begin{acknowledgement}
~\\
We thank L.Dixon and D.Kosower for stimulating discussions, and S.Hoeche for providing data from BlackHat.\par

We thank CERN for the very successful operation of the LHC, as well as the support staff from our institutions without whom ATLAS could not be
operated efficiently.

We acknowledge the support of ANPCyT, Argentina; YerPhI, Armenia; ARC, Australia; BMWF, Austria; ANAS, Azerbaijan; SSTC, Belarus; CNPq and FAPESP,
Brazil; NSERC, NRC and CFI, Canada; CERN; CONICYT, Chile; CAS, MOST and NSFC, China; COLCIENCIAS, Colombia; MSMT CR, MPO CR and VSC CR, Czech Republic;
DNRF, DNSRC and Lundbeck Foundation, Denmark; EPLANET and ERC, European Union; IN2P3-CNRS, CEA-DSM/IRFU, France; GNAS, Georgia; BMBF, DFG, HGF, MPG 
and AvH Foundation, Germany; GSRT, Greece; ISF, MINERVA, GIF, DIP and Benoziyo Center, Israel; INFN, Italy; MEXT and JSPS, Japan; CNRST, Morocco; FOM 
and NWO, Netherlands; RCN, Norway; MNiSW, Poland; GRICES and FCT, Portugal; MERYS (MECTS), Romania; MES of Russia and ROSATOM, Russian Federation; 
JINR; MSTD, Serbia; MSSR, Slovakia; ARRS and MVZT, Slovenia; DST/NRF, South Africa; MICINN, Spain; SRC and Wallenberg Foundation, Sweden; SER, SNSF 
and Cantons of Bern and Geneva, Switzerland; NSC, Taiwan; TAEK, Turkey; STFC, the Royal Society and Leverhulme Trust, United Kingdom; DOE and NSF, 
United States of America.

The crucial computing support from all WLCG partners is acknowledged gratefully, in particular from CERN and the ATLAS Tier-1 facilities at
TRIUMF (Canada), NDGF (Denmark, Norway, Sweden), CC-IN2P3 (France), KIT/GridKA (Germany), INFN-CNAF (Italy), NL-T1 (Netherlands), PIC (Spain),
ASGC (Taiwan), RAL (UK) and BNL (USA) and in the Tier-2 facilities worldwide.

\end{acknowledgement}

\bibliographystyle{atlasnote} \bibliography{WHelicityPaper}

\clearpage
\onecolumn
\begin{flushleft}
{\Large The ATLAS Collaboration}

\bigskip

G.~Aad$^{\rm 48}$,
B.~Abbott$^{\rm 110}$,
J.~Abdallah$^{\rm 11}$,
A.A.~Abdelalim$^{\rm 49}$,
A.~Abdesselam$^{\rm 117}$,
O.~Abdinov$^{\rm 10}$,
B.~Abi$^{\rm 111}$,
M.~Abolins$^{\rm 87}$,
O.S.~AbouZeid$^{\rm 157}$,
H.~Abramowicz$^{\rm 152}$
% ...
\bigskip

% ...
$^{10}$ Institute of Physics, Azerbaijan Academy of Sciences, Baku, Azerbaijan\\
$^{11}$ Institut de F\'isica d'Altes Energies and Departament de F\'isica de la Universitat Aut\`onoma  de Barcelona and ICREA, Barcelona, Spain\\
% ...
$^{48}$ Fakult\"{a}t f\"{u}r Mathematik und Physik, Albert-Ludwigs-Universit\"{a}t, Freiburg i.Br., Germany\\
$^{49}$ Section de Physique, Universit\'e de Gen\`eve, Geneva, Switzerland\\
% ...
$^{87}$ Department of Physics and Astronomy, Michigan State University, East Lansing MI, United States of America\\
% ...
$^{110}$ Homer L. Dodge Department of Physics and Astronomy, University of Oklahoma, Norman OK, United States of America\\
$^{111}$ Department of Physics, Oklahoma State University, Stillwater OK, United States of America\\
% ...
$^{117}$ Department of Physics, Oxford University, Oxford, United Kingdom\\
% ...
$^{152}$ Raymond and Beverly Sackler School of Physics and Astronomy, Tel Aviv University, Tel Aviv, Israel\\
% ...
$^{157}$ Department of Physics, University of Toronto, Toronto ON, Canada\\
% ...
\end{flushleft}

\end{document}